\documentclass[preprint,aps,superscriptaddress,nofootinbib]{revtex4-1}
\usepackage[T1]{fontenc}
\usepackage[latin9]{inputenc}
\usepackage{color}
\usepackage{array}
\usepackage{rotfloat}
\usepackage{units}
\usepackage{amsmath}
\usepackage{amssymb}
\usepackage{graphicx}
\usepackage{esint}
\usepackage[unicode=true,pdfusetitle,
 bookmarks=true,bookmarksnumbered=false,bookmarksopen=false,
 breaklinks=false,pdfborder={0 0 1},backref=false,colorlinks=true]
 {hyperref}

\makeatletter

\providecommand{\tabularnewline}{\\}

 \@ifundefined{textcolor}{}
 {%
   \definecolor{BLACK}{gray}{0}
   \definecolor{WHITE}{gray}{1}
   \definecolor{RED}{rgb}{1,0,0}
   \definecolor{GREEN}{rgb}{0,1,0}
   \definecolor{BLUE}{rgb}{0,0,1}
   \definecolor{CYAN}{cmyk}{1,0,0,0}
   \definecolor{MAGENTA}{cmyk}{0,1,0,0}
   \definecolor{YELLOW}{cmyk}{0,0,1,0}
 }

\usepackage{array}
\usepackage{units}
\usepackage{bm}
\usepackage{braket}
\usepackage{slashed}
\usepackage[bbgreekl]{mathbbol}
\usepackage{bbm}

\makeatother

\begin{document}

\title{Constraining neutrino electromagnetic properties by germanium detectors}

\author{Jiunn-Wei Chen}

\email{jwc@phys.ntu.edu.tw}

\affiliation{Department of Physics, National Taiwan University, Taipei 10617,
Taiwan}

\affiliation{National Center for Theoretical Sciences and Leung Center for Cosmology
and Particle Astrophysics, National Taiwan University, Taipei 10617,
Taiwan}

\author{Hsin-Chang Chi}

\email{hsinchang@mail.ndhu.edu.tw}

\affiliation{Department of Physics, National Dong Hwa University, Shoufeng, Hualien
97401, Taiwan}

\author{Keh-Ning Huang}

\affiliation{Department of Physics, National Taiwan University, Taipei 10617,
Taiwan}

\affiliation{Department of Physics, Sichuan University, Chengdu, Sichuan, China}

\affiliation{Department of Physics, Fuzhou University, Fuzhou, Fujian, China}

\author{Hau-Bin Li}

\affiliation{Institute of Physics, Academia Sinica, Taipei 11529, Taiwan}

\author{C.-P. Liu}

\email{cpliu@mail.ndhu.edu.tw}

\affiliation{Department of Physics, National Dong Hwa University, Shoufeng, Hualien
97401, Taiwan}

\author{Lakhwinder Singh}

\affiliation{Institute of Physics, Academia Sinica, Taipei 11529, Taiwan}

\affiliation{Department of Physics, Banaras Hindu University, Varanasi 221005,
India}

\author{Henry T. Wong}

\affiliation{Institute of Physics, Academia Sinica, Taipei 11529, Taiwan}

\author{Chih-Liang Wu}

\affiliation{Department of Physics, National Taiwan University, Taipei 10617,
Taiwan}

\affiliation{Institute of Physics, Academia Sinica, Taipei 11529, Taiwan}

\author{Chih-Pan Wu}

\affiliation{Department of Physics, National Taiwan University, Taipei 10617,
Taiwan}
\begin{abstract}
The electromagnetic properties of neutrinos, which are either trivial
or negligible in the context of the Standard Model, can probe new
physics and have significant implications in astrophysics and cosmology.
The current best direct limits on the neutrino millicharges and magnetic
moments are both derived from data taken with germanium detectors
with low thresholds at keV levels. In this paper, we discuss in detail
a robust, \textit{ab initio} method: the multiconfiguration relativistic
random phase approximation, that enables us to reliably understand
the germanium detector response at the sub-keV level, where atomic
many-body physics matters. Using existing data with sub-keV thresholds,
limits on reactor antineutrino's millicharge, magnetic moment, and
charge radius squared are derived. The projected sensitivities for
next generation experiments are also given and discussed.
\end{abstract}
\maketitle

\section{Introduction}

Investigations of neutrino properties continue to be an accretive
field of emerging interests to both theoretical and experimental physicists.
Their nonzero masses, as suggested by neutrino oscillation experiments
with various sources, already hint the necessity of extending the
Standard Model (SM) to accommodate massive neutrinos. It is no wonder
that their properties such as absolute masses, mass hierarchy, Dirac
or Majorana nature, and precise mixing parameters are among the most
actively pursued topics in neutrino physics for their great discovery
potential. 

Another interesting venue to look for surprises in neutrinos is their
nontrivial electromagnetic (EM) properties (see, e.g., \cite{Giunti:2014ixa,*Broggini:2012df,*Wong:2005pa}
for recent reviews). In the SM, neutrinos are strictly neutral. Their
tiny charge radii squared, magnetic dipole moments, anapole moments
(require parity violation in addition), and electric dipole moments
(require both parity and time-reversal violation in addition) only
arise in forms of radiative corrections (in some cases, finite mass
terms and flavor mixing matrix have to be included). Going beyond
the SM, there are numerous conjectures of larger neutrino EM moments,
including neutrinos being millicharged. The present best upper limits
on some of these moments, either set directly by experiments, or inferred
indirectly from observational evidences combined with theoretical
arguments, are orders of magnitude larger than the SM predictions
(see \cite{Agashe:2014kda} and references therein for the current
status). As a result, this leaves space for new physics. Also, the
additional EM interactions with the copious amount of neutrinos in
the universe will have significant implications for astrophysics and
cosmology. 

It was recently identified \cite{Wong:2010pb,Chen:2014dsa} that the
unexplored interaction channel of neutrino-induced atomic ionization:
\[
\nu+\mathrm{A}\rightarrow\nu+\mathrm{A}^{-}+e^{-}\,,
\]
is an interesting avenue to study possible neutrino electromagnetic
effects, and has the potentials of producing surprises. The germanium
atom (Ge) is selected for the studies, since there are matured Ge
detector techniques with low (at the atomic transition range of keV)
threshold and good resolution to resolve possible spectra structures
and peaks and end-points, which are essential to provide smoking-gun
positive signatures. Existing data from the TEXONO and GEMMA experiments
with reactor neutrinos already provide bounds on neutrino magnetic
moments~\cite{Li:2002pn,Wong:2006nx,Beda:2012zz,Beda:2013mta}, neutrino
charge radius~\cite{Deniz:2010mp}, and milli-charges~\cite{Gninenko:2006fi,Studenikin:2013my}.
New generations of Ge detectors capable of measuring events as low
as 100 eV are expected to further expand the sensitivities~\cite{Lin:2007ka,Li:2013fla,Zhao:2013xsf,Agnese:2014aze}. 

To interpret experimental data and put limits on these moments, an
important theoretical input: the differential cross section formulae
for neutrino scattering in detectors, is necessary (see, e.g., Ref.~\cite{Kouzakov:2014lka}
for a recent review of neutrino-atom collision theory). While the
conventional approach of treating the atomic electrons as free particles
is considered a good approximation at high energies, at sub-keV regime,
which is similar to atomic scales, proper treatments of many-electron
dynamics in atomic ionization must be incorporated for a better understanding
of detector responses at low energies. 

Motivated by this goal, we recently applied \textsl{ab initio} calculations
in the framework of multiconfiguration relativistic random phase approximation
(MCRRPA) theory to study the atomic ionization of germanium by neutrino
scattering. Partial results were reported in \cite{Chen:2013lba}
and \cite{Chen:2014dsa}, which dealt with the neutrino magnetic moment
and millicharge, respectively. 

The purpose of this article is twofold: On the theory part, we present
our approach in full details, elaborate in particular the benchmark
calculations that serve as a concrete basis on which the method and
uncertainty estimate can be justified, and consider all observables
that can be probed by atomic ionization. Comparisons with previous
works~ \cite{Fayans:1992kk,Kopeikin:1997ge,Fayans:2001pg,Kopeikin:2003bx,Wong:2010pb,Voloshin:2010vm,Kouzakov:2010tx,Kouzakov:2011vx}
are given so that differences in various approaches and the applicability
of various approximation schemes at the sub-keV regime can be clearly
examined.

The organization of this paper is as follows. In Section \ref{sec:formulatiion},
we give the general formulation of atomic ionization by neutrinos,
and mention two widely-used approximation schemes: free electron approximation
and equivalent photon approximation in \ref{sub:FEA} and \ref{sub:EPA},
respectively. Our approach to atomic many-body problems: the multiconfiguration
relativistic random phase approximation, is outlined in Section \ref{sub:MCRRPA},
and its application to the structure and photoionization of germanium
atoms are described in Section \ref{sub:Ge_MCDF} and \ref{sub:Ge_photo},
subsequently. In Section \ref{sub:Ge_nuAI}, we present and discuss
our results for germanium ionization by neutrino scattering, and compare
with existing works. Limits on neutrino electromagnetic moments are
derived in Section \ref{sub:TEXONO} by using realistic reactor antineutrino
spectra and data. As there have been proposals of using neutrinos
from tritium $\beta$ decay to study neutrino magnetic moment~\cite{Kopeikin:2003bx,Giomataris:2003bp,McLaughlin:2003yg},
our calculation for this case is presented in Section \ref{sub:3H_beta}.
The summary is in Section \ref{sec:summary}, and the technical details
of multipole expansion, which is relevant to our calculations, is
in Appendix~\ref{sec:app-1}.

\section{Formulation of Atomic Ionization by Neutrinos \label{sec:formulatiion}}

Consider the ionization of an atom A by scattering a neutrino $\nu_{l}$
($l$ denoting the flavor eigenstate) off atomic bound electrons 
\begin{equation}
\nu_{l}+\mathrm{A}\rightarrow\nu_{l}+\mathrm{A}^{+}+e^{-}\,.
\end{equation}
For $l=\mu,\tau$, the process only proceeds through the neutral weak
interaction (in $t$-channel), while for $l=e$, the charged weak
interaction (in $s$-channel) also contributes. Using a Feirz reordering,
the general low-energy weak scattering amplitude can be compactly
gathered in one formula 
\begin{equation}
\mathcal{M}^{(w)}=\frac{G_{F}}{\sqrt{2}}j_{\mu}^{(w)}(c_{V}\mathcal{J}^{\mu}-c_{A}\mathcal{J}_{5}^{\mu})\,,\label{eq:amp_weak}
\end{equation}
where $G_{F}$ is the Fermi constant. The neutrino weak current 
\begin{equation}
j_{\mu}^{(w)}=\bar{\nu}(k_{2},s_{2})\gamma_{\mu}(1-\gamma_{5})\nu(k_{1},s_{1})\,,\label{eq:j_nu_weak}
\end{equation}
takes on the usual Dirac bilinear form with $k_{1}=(\omega_{1},\vec{k_{1}})$,
$k_{2}=(\omega_{2},\vec{k_{2}})$ being the four-momenta and $s_{1}$,
$s_{2}$ being the helicity states of the neutrino before and after
scattering, respectively. The energy and 3-momentum transfer by the
neutrinos are defined as 
\begin{equation}
q^{\mu}=(T,\vec{q})=(\omega_{1}-\omega_{2},\vec{k}_{1}-\vec{k}_{2})\,.
\end{equation}
The atomic (axial-)vector current, $\mathcal{J}_{(5)}^{\mu}$, 
\begin{equation}
\mathcal{J}_{(5)}^{\mu}\equiv\braket{\Psi_{f}|\hat{\mathcal{J}}_{(5)}^{\mu}(-\vec{q})|\Psi_{i}}=\int d^{3}x\, e^{i\vec{q}\cdot\vec{x}}\braket{\Psi_{f}|\hat{\bar{\psi}}_{e}(\vec{x})\gamma^{\mu}(\gamma_{5})\hat{\psi}_{e}(\vec{x})|\Psi_{i}}\,,\label{eq:J_atom}
\end{equation}
is the matrix element of a one-electron (axial-)vector current operator
$\hat{\mathcal{J}}_{(5)}^{\mu}(-\vec{q})$ (in momentum space) evaluated
with many-body atomic initial and final states, $\ket{\Psi_{i}}$
and $\ket{\Psi_{f}}$. The vector and axial-vector coupling constants
are 
\begin{align}
c_{V}=-\frac{1}{2}+2\sin^{2}\theta_{w}+\delta_{l,e}\,, & \qquad c_{A}=-\frac{1}{2}+\delta_{l,e}\,,
\end{align}
where $\theta_{w}$ is the Weinberg angle. The extra Kronecker delta
is added to account for the additional $s$-channel scattering for
$\nu_{e}$.

Now suppose a neutrino has nonzero electromagnetic (EM) moments; in
the most general case, the associated EM current can be expressed
as 
\begin{align}
j_{\mu}^{(\gamma)}= & \bar{\nu}(k_{2},s_{2})\left[F_{1}(q^{2})\gamma_{\mu}-i(F_{2}(q^{2})+iF_{E}(q^{2})\gamma_{5})\sigma_{\mu\nu}q^{\nu}+F_{A}(q^{2})(q^{2}\gamma_{\mu}-\slashed{q}q_{\mu})\gamma_{5}\right]\nu(k_{1},s_{1})\label{eq:j_nu_EM}
\end{align}
where $q^{2}\equiv q_{\mu}q^{\mu}$. The four terms $F_{1}(q^{2})$,
$F_{2}(q^{2})$, $F_{A}(q^{2})$, and $F_{E}(q^{2})$ are referred
as the charge, anomalous magnetic, anapole, and electric dipole form
factors, respectively. Up to the order of $q^{2}$ in $j_{\mu}^{(\gamma)}$,
we define the electric charge, charge radius squared, magnetic dipole
moment, anapole moment, and electric dipole moment of a neutrino by
\begin{align}
\mathbbm{q}_{\nu} & =F_{1}(0)\,,\nonumber \\
\braket{\mathbbm{r}_{\nu}^{2}} & =\left.6\frac{d}{dq^{2}}F_{1}(q^{2})\right|_{q^{2}\rightarrow0}\,,\nonumber \\
\bbkappa_{\nu} & =F_{2}(0)\,,\nonumber \\
\mathbbm{a}_{\nu} & =F_{A}(0)\,,\nonumber \\
\mathbbm{d}_{\nu} & =F_{E}(0)\,,
\end{align}
respectively, and they are all measured in the fundamental charge
units $e$. Note that the existence of $\mathbbm{a}_{\nu}$ violates
parity conservation, and $\mathbbm{d}_{\nu}$ violates both parity
and time-reversal conservation. Also, in the Standard Model, the values
of both $\braket{\mathbbm{r}_{\nu}^{2}}$ and $\mathbbm{a}_{\nu}$
arising from electroweak radiative corrections are not gauge-independent
quantities; only after the full radiative corrections being considered
are the gauge-independent, physical observables resulted in~\cite{Musolf:1990sa}.
While there are attempts to define these moments in gauge-independent
manners, but still controversial. Here we do not concern ourselves
further with such subtleties, but just practically assume these exotic
moments, whose definitions are consistent with current conservation
as obviously seen in Eq.~(\ref{eq:j_nu_EM}), exist, and study their
contributions in scattering processes.

Any nonzero EM moments of a neutrino therefore generate additional
contributions to the atomic ionization process; they are given by
the associated EM scattering amplitude~%
\footnote{We note that a non-zero $\mathbbm{q}_{\nu}$ also induces extra neutral
weak interactions which modify Eq.~\ref{eq:amp_weak} at the order
of $\mathbbm{q}_{\nu}\sin^{2}\theta_{w}$; therefore can be safely
ignored.%
} 
\begin{equation}
\mathcal{M}^{(\gamma)}=\frac{4\pi\alpha}{q^{2}}j_{\mu}^{(\gamma)}\mathcal{J}^{\mu}\,.\label{eq:amp_EM}
\end{equation}
Before presenting the complete scattering formula, we discuss a few
kinematical considerations that help to reduce the full result to
a simpler form.

First, as neutrinos are much lighter than all the energy scales relevant
to the atomic ionization processes of concern, an ultrarelativistic
limit $m_{\nu}\rightarrow0$ is considered a good approximation. In
such cases, the chirality and helicity states of a neutrino are the
same, so scattering amplitudes of neutrino-helicity-flipping interactions
with $\bbkappa_{\nu}$ and $\mathbbm{d}_{\nu}$, do not interfere
with ones of neutrino-helicity-conserving interactions. On the other
hand, since weak interactions and those with $\mathbbm{q}_{\nu}$,
$\braket{\mathbbm{r}_{\nu}^{2}}$, and $\mathbbm{a}_{\nu}$ all preserve
helicity, there are interference terms between the weak and EM amplitudes.
Their magnitudes are important when constraints of $\mathbbm{q}_{\nu}$,
$\braket{\mathbbm{r}_{\nu}^{2}}$, and $\mathbbm{a}_{\nu}$ are to
be extracted from experimental data.

Second, the interaction with $\braket{\mathbbm{r}_{\nu}^{2}}$ apparently
takes a four-Fermi contact form (evidenced by the $1/q^{2}$ photon
propagator being cancelled by the $q^{2}$ factor in the associated
current), and so does the interaction with $\mathbbm{a}_{\nu}$~\cite{Musolf:1990sa}.
As a result, the combined EM scattering amplitude 
\begin{equation}
\mathcal{M}^{(\braket{\mathbbm{r}_{\nu}^{2}}+\mathbbm{a}_{\nu})}=4\pi\alpha[\bar{\nu}\gamma_{\mu}(\nicefrac{1}{6}\braket{\mathbbm{r}_{\nu}^{2}}+\mathbbm{a}_{\nu}\gamma_{5})\nu]\mathcal{J}^{\mu}\,,
\end{equation}
look similar to $\mathcal{M}^{(w)}$, except no coupling to the atomic
axial-vector current $\mathcal{J}_{5}^{\mu}$.

Third, by the identities 
\begin{equation}
\bar{\nu}_{L}\gamma_{\mu}\nu_{L}=-\bar{\nu}_{L}\gamma_{\mu}\gamma_{5}\nu_{L}\,,\quad\bar{\nu}_{R}\sigma_{\mu\nu}\nu_{L}=-\bar{\nu}_{R}\sigma_{\mu\nu}\gamma_{5}\nu_{L}\,,
\end{equation}
one deduces that $\braket{\gamma_{\nu}^{2}}$ and $a_{\nu}$ can not
be distinguished in ultrarelativistic neutrino scattering and should
effectively appear as one moment, the effective charge radius squared:
\begin{equation}
\braket{\mathbbm{r}_{\nu}^{2}}^{(\mathrm{eff})}=\braket{\mathbbm{r}_{\nu}^{2}}-6\mathbbm{a}_{\nu}\,.
\end{equation}
The same argument applies to $\kappa_{\nu}$ and $d_{\nu}$ that they
appear as one effective anomalous magnetic moment: 
\begin{equation}
\bbkappa_{\nu}^{(\mathrm{eff})}=\bbkappa_{\nu}-i\mathbbm{d}_{\nu}\,.
\end{equation}

Starting from the total scattering amplitude, $\mathcal{M}^{(w)}+\mathcal{M}^{(\gamma)}$,
and following the standard procedure, the single differential cross
section with respect to neutrino energy deposit $T$ for an inclusive
process with a unpolarized target is obtained. When there is only
weak scattering, the result is 
\begin{eqnarray}
\dfrac{d\sigma^{(w)}}{dT} & = & \dfrac{G_{F}^{2}}{\pi}(E_{\nu}-T)^{2}\int d\cos\theta\,\cos^{2}\dfrac{\theta}{2}\Big\{ R_{00}^{(w)}-\dfrac{T}{|\vec{q}|}R_{03+30}^{(w)}+\dfrac{T^{2}}{|\vec{q}|^{2}}R_{33}^{(w)}\nonumber \\
 & + & \left(\tan^{2}\dfrac{\theta}{2}-\dfrac{q^{2}}{2|\vec{q}|^{2}}\right)R_{11+22}^{(w)}+\tan\dfrac{\theta}{2}\sqrt{\tan^{2}\dfrac{\theta}{2}-\dfrac{q^{2}}{|\vec{q}|^{2}}}R_{12+21}^{(w)}\Big\}\label{eq:dS/dT_weak}
\end{eqnarray}
where $\theta$ is the neutrino scattering angle, $E_{\nu}$ is the
incident neutrino energy. The atomic weak response functions 
\begin{eqnarray}
R_{\mu\nu}^{(w)} & = & \frac{1}{2J_{i}+1}\sum_{M_{J_{i}}}\sum_{f}\braket{\Psi_{f}|c_{V}\hat{\mathcal{J}}_{\mu}-c_{A}\hat{\mathcal{J}}_{5\mu}|\Psi_{i}}\braket{\Psi_{f}|c_{V}\hat{\mathcal{J}}_{\nu}-c_{A}\hat{\mathcal{J}}_{5\nu}|\Psi_{i}}^{*}\nonumber \\
 &  & \times\delta(T+E_{i}-E_{f})\,,\label{eq:RF_weak}
\end{eqnarray}
involve a sum of the final scattering states $\ket{\Psi_{f}}$ and
a spin average of the initial states $\ket{\Psi_{i}}=\ket{J_{i},M_{J_{i}},\ldots}$,
and the Dirac delta function imposes energy conservation. The Greek
indices $\mu,\nu$ take values $0,1,2,3$, and without loss of generality,
the direction of $\vec{q}$ is taken to be the quantization axis with
$\mu=3$. 

The contributions from the helicity-conserving (h.c.) interactions
with $\mathbbm{q}_{\nu}$ and $\braket{\mathbbm{r}_{\nu}}^{\mathrm{(eff)}}$,
as they interfere with the weak scattering, can be compactly included
by the following substitution 
\begin{align}
\frac{d\sigma^{(w)}}{dT}\rightarrow\frac{d\sigma^{(\mathrm{h.c.})}}{dT}\,, & \quad\mathrm{with}\, c_{V}\rightarrow c_{V}+2\sqrt{2}\pi\frac{\alpha}{G_{F}}(\frac{1}{q^{2}}\mathbbm{q}_{\nu}^{2}+\frac{1}{6}\braket{\mathbbm{r}_{\nu}^{2}}^{(\mathrm{eff})})\,.\label{eq:dS/dT_h.c.}
\end{align}
It should be pointed out that the inclusion of $\mathbbm{q}_{\nu}$
is only formally, as it goes with a kinematics-dependent term $1/q^{2}$
that differentiates its contribution from the other contact interactions.

As will be explicitly shown later, the contribution from $\mathbbm{q}_{\nu}$
with the current upper limit $\lesssim10^{-12}$ derived from direct
measurements dominates over the weak scattering. When the $\mathbbm{q}_{\nu}$-weak
interference terms are much less important, it is convenient to isolate
the pure Coulomb (coul) scattering part,

\begin{eqnarray}
\dfrac{d\sigma^{(\mathrm{coul})}}{dT} & =\mathbbm{q}_{\nu}^{2} & (2\pi\alpha^{2})\left(1-\frac{T}{E_{\nu}}\right)\int d\cos\theta\,\Big\{\dfrac{(2E_{\nu}-T)^{2}-|\vec{q}|^{2}}{|\vec{q}|{}^{4}}R_{00}^{(\gamma)}\nonumber \\
 & - & \left[\dfrac{q^{2}+4E_{\nu}(E_{\nu}-T)}{2|\vec{q}|^{2}q^{2}}+\dfrac{1}{q^{2}}\right]R_{11+22}^{(\gamma)}\Big\}\,,\label{eq:dS/dT_F1}
\end{eqnarray}
which is proportional to $\mathbbm{q}_{\nu}^{2}$. In such cases,
we apply the approximated form 
\begin{equation}
\left.\frac{d\sigma^{(\mathrm{h.c.})}}{dT}\right|_{\mathrm{large\,}\mathbbm{q}_{\nu}}\approx\left.\frac{d\sigma^{(\mathrm{h.c.})}}{dT}\right|_{c_{V}\rightarrow c_{V}+\frac{\sqrt{2}\pi\alpha}{3G_{F}}\braket{\mathbbm{r}_{\nu}^{2}}^{(\mathrm{eff})}}+\frac{d\sigma^{(\mathrm{coul})}}{dT}\,.
\end{equation}

On the other hand, the contribution from the the helicity-violating
(h.v.) interaction with $\bbkappa_{\nu}^{(\mathrm{eff})}$ has no
interference with the helicity-conserving part so that 
\begin{align}
\dfrac{d\sigma}{dT} & =\dfrac{d\sigma^{(\mathrm{h.c.})}}{dT}+\dfrac{d\sigma^{(\mathrm{h.v.})}}{dT}\,,\label{eq:dS/dT_full}
\end{align}
with

\begin{eqnarray}
\dfrac{d\sigma^{(\mathrm{h.v.})}}{dT} & = & (\bbkappa_{\nu}^{2}+\mathbbm{d}_{\nu}^{2})(2\pi\alpha^{2})\left(1-\frac{T}{E_{\nu}}\right)\int d\cos\theta\,\Big\{-\dfrac{(2E_{\nu}-T)^{2}q^{2}}{|\vec{q}|{}^{4}}\, R_{00}^{(\gamma)}\nonumber \\
 &  & +\dfrac{q^{2}+4E_{\nu}(E_{\nu}-T)}{2|\vec{q}|^{2}}R_{11+22}^{(\gamma)}\Big\}\,.\label{eq:dS/dT_F2}
\end{eqnarray}
Note that the EM response functions appearing in Eqs.~(\ref{eq:dS/dT_F1})
and (\ref{eq:dS/dT_F2}) are related to the weak response functions
by setting $c_{V}=1$ and $c_{A}=0$ in Eq.~(\ref{eq:RF_weak}) 
\begin{equation}
\left.R_{\mu\nu}^{(w)}\right|_{c_{V}=1,c_{A}=0}\rightarrow R_{\mu\nu}^{(\gamma)}\,,\label{eq:RF_EM}
\end{equation}
as EM interactions only couple to vector currents (which result in
$R_{12+21}^{(\gamma)}=0$). Because of vector current conservation,
the longitudinal part of a spatial current density $(\mu=3)$ is related
to the charge density ($\mu=0$). Therefore the response functions
$R_{03+30}^{(\gamma)}$ and $R_{33}^{(\gamma)}$ are subsumed in $R_{00}^{(\gamma)}$
.

A couple of important remarks on kinematics in $\nicefrac{d\sigma}{dT}$
are due here: (i) For fixed $E_{\nu}$ and $T$, the square of four
momentum transfer $q^{2}$ in the ultrarelativistic limit is determined
by the neutrino scattering angle $\theta$ 
\begin{equation}
q^{2}=-4E_{\nu}^{2}\sin^{2}(\frac{\theta}{2})-m_{\nu}^{2}\frac{T^{2}}{E_{\nu}^{2}}\,.\label{eq:qsquare}
\end{equation}
It will not vanish even at the forward angle $\theta=0$ as long as
the neutrino is not massless $m_{\nu}\neq0$. (This is important for
scattering with $\mathbbm{q}_{\nu}$). (ii) By four momentum conservation,
the integration variable $\cos\theta$ is constrained by 
\begin{align}
\min\left\{ 1,\max\left[-1,\dfrac{E_{\nu}^{2}+(E_{\nu}-T)^{2}-2M_{\mathrm{A}}(T-B)}{2E_{\nu}(E_{\nu}-T)}\right]\right\} \leq\cos\theta\leq1\,,\label{eq:cos_range}
\end{align}
where $M_{\mathrm{A}}$ is the atomic mass and $B$ is the binding
energy of the ejected electron.

To evaluate $\nicefrac{d\sigma}{dT}$, the most challenging task is
the calculation of all relevant atomic response functions, Eqs.(\ref{eq:RF_weak},\ref{eq:RF_EM}).
Before discussing our \textit{ab initio} approach in next section,
we review a couple of simple approximation schemes that work in certain
kinematic regimes and by which tedious many-body calculations can
be spared.

\subsection{Free Electron Approximation \label{sub:FEA}}

In case of high energy scattering when electron binding energy is
comparatively negligible, a conceptually straightforward approach
is to use a neutrino-free-electron scattering formula $\nicefrac{d\sigma^{(0)}}{dT}$.
The number of atomic electrons can be freed depends on the neutrino
energy deposition $T$. By introducing a step function $\theta(T-B_{i})$
to judge whether the $i$th electron, with binding energy $B_{i}$,
can contribute to the scattering process, one obtains the conventional
scattering formula based on the free electron approximation

\begin{equation}
\left.\dfrac{d\sigma}{dT}\right|_{\mathrm{FEA}}=\sum_{i=1}^{Z}\theta(T-B_{i})\left.\dfrac{d\sigma^{(0)}}{dT}\right|_{q^{2}=-2m_{e}T}\,.\label{eq:dS/dT_FEA}
\end{equation}

Despite FEA enjoys a lot of success in many situations, its applicability
is not always self-evident, in particular when issues like relevant
energy scales and kinematics of concern arise. For example, as it
was shown explicitly in Ref. for hydrogen-like atoms: (i) The borderline
incident neutrino energy above which FEA can apply is the binding
momentum $\sim Zm_{e}\alpha$, instead of the binding energy $\sim Z^{2}m_{e}\alpha^{2}$.
(ii) Because FEA only has a specific $q^{2}=-2m_{e}T$ in contrary
to an allowed range prescribed by Eqs.~(\ref{eq:qsquare},\ref{eq:cos_range})
for the realistic case, it fails to be valid for relativistic muon
scattering and nonrelativistic WIMP scattering. Therefore, to reduce
the potential errors caused this conventional practice in particular
for detector's response at low energies is an important theoretical
task.

\subsection{Equivalent Photon Approximation \label{sub:EPA}}

In typical EM scattering with ultrarelativistic charged particles,
it was long established that the equivalent photon approximation (EPA)
is well founded~\cite{Weizsacker:1934sx,Williams:1934ad,Greiner:QED}.
Such processes mostly happen with peripheral scattering angles, i.e.,
$q^{2}\rightarrow0$; it is thus obvious from Eq.~(\ref{eq:dS/dT_F1})
that the contribution from the transverse response function $R_{11+22}^{(\gamma)}$
dominates and the longitudinal part $R_{00}^{(\gamma)}$ can be ignored.
As the ``on-shell'' transverse response function is directly linked
to the total cross section of photoabsorption

\begin{align}
\sigma_{\mathrm{abs}}^{(\gamma)}(T) & =\dfrac{2\pi^{2}\alpha}{T}R_{11+22}^{(\gamma)}(q^{2}=0)\,,\label{eq:S_abs}
\end{align}
the EPA further approximates $R_{11+22}^{(\gamma)}(q^{2})\approx R_{11+22}^{(\gamma)}(0)$
so that the Coulomb differential cross section for $\mathbbm{q}_{\nu}$
\begin{align}
\left.\dfrac{d\sigma^{(\mathrm{coul})}}{dT}\right|_{\mathrm{EPA}} & =-\mathbbm{q}_{\nu}^{2}(\frac{\alpha}{\pi})(1-\frac{T}{E_{\nu}})\sigma_{\mathrm{abs}}^{(\gamma)}(T)\int\, d\cos\theta\left[\dfrac{q^{2}+4E_{\nu}(E_{\nu}-T)}{2|\vec{q}|^{2}q^{2}}+\dfrac{1}{q^{2}}\right]\,,\label{eq:dS/dT_F1_EPA}
\end{align}
can be directly determined by experiment.

Applying similar procedure to EM scattering with $\bbkappa_{\nu}$
and $\mathbbm{d}_{\nu}$:

\begin{align}
\left.\dfrac{d\sigma^{(\mathrm{h.v.})}}{dT}\right|_{\mathrm{EPA}} & =(\bbkappa_{\nu}^{2}+\mathbbm{d}_{\nu}^{2})(\frac{\alpha}{\pi})(1-\frac{T}{E_{\nu}})\sigma_{\mathrm{abs}}^{(\gamma)}(T)\int\, d\cos\theta\left[\dfrac{q^{2}+4E_{\nu}(E_{\nu}-T)}{2|\vec{q}|^{2}}\right]\,,\label{eq:dS/dT_F2_EPA}
\end{align}
the cross section formula differs noticeably from the previous case
by missing of the $1/q^{2}$ enhancement in the real photon limit.
For more discussions about why there should not be atomic-enhanced
sensitivities to neutrino magnetic moments at low energies, in contrary
to what was claimed in Ref.~\cite{Wong:2010pb} (which is based on
a slight different twist of Eq.~\ref{eq:dS/dT_F2_EPA}), we refer
readers to Refs.~\cite{Voloshin:2010vm,Chen:2013iud} for details.

\section{\textit{ab initio} Description of Germanium}

To go beyond the simple approximation schemes mentioned in the last
section and evaluate the cross section formulae more reliably at low
energies, the structure and ionization of detector atoms have to be
considered on a more elaborate basis. In this section, we first introduce
our approach to the atomic many-body problems: the multiconfiguration
relativistic random phase approximation (MCRRPA) theory. In the following
subsections, we present our results for the structure and photoionization
of germanium atoms, respectively, and benchmark the quality of MCRRPA
as reliable approach to describe the responses of germanium detectors.

\subsection{The MCRRPA Theory \label{sub:MCRRPA}}

The relativistic random-phase approximation (RRPA) has been applied,
with remarkable successes, to photoexcitation and photoionization
of closed-shell atoms and ions of high nuclear charge, such as heavy
noble gas atoms, where the ground state is well isolated from the
excited states. For other closed-shell systems, such as alkaline-earth
atoms, which have low-lying excited states, such applications have
been less successful, owing to the importance of two-electron excitations
which are omitted in the RRPA. The MCRRPA theory is a generalization
RRPA by using a multiconfiguration wave function as the reference
state which is suitable for treating photoexcitation and photoionization
of closed-shell and certain open-shell systems of high nuclear charge.
The great success it achieved in various atomic radiative processes
can be found in Refs.~\cite{Huang:1995cc}. A detailed formulation
of the MCRRPA has been given in a previous paper~\cite{Huang:1981wj},
and we summarize the essential features here.

One way to derive the MCRRPA equations is through linearizing the
time-dependent multiconfiguration Hartree-Fock (TDHF) equations.~%
\footnote{An alternatively derivation from an equation-of-motion point of view
is given in Ref.~\cite{Huang:1982re}.%
} For a \textit{N}-electron atomic system, the time-dependent relativistic
Hamiltonian is given by 
\begin{equation}
H(t)=H_{0}+V(t)\,,
\end{equation}
where the unperturbed Hamiltonian 
\begin{equation}
H_{0}={\displaystyle \sum_{i=1}^{N}h(\vec{r}_{i})+{\displaystyle \sum_{i<j}^{N}\dfrac{e^{2}}{r_{ij}}\,,}}
\end{equation}
contains the sum of single-electron Dirac Hamiltonians

\begin{equation}
h(\vec{r})=c\vec{\alpha}\cdot\vec{p}+\beta c^{2}-\frac{Z}{r}\,,
\end{equation}

\noindent and the Coulomb repulsion between two-electron pairs (the
latter summation); and the time-dependent external perturbation

\begin{equation}
V(t)=\sum_{i=1}^{N}v_{+}(\vec{r}_{i})e^{-i\omega t}+\sum_{i=1}^{N}v_{-}(\vec{r}_{i})e^{+i\omega t}\,,
\end{equation}
takes a harmonic form and induces transitions between atomic states.
Note that atomic units (a.u.) are employed throughout this paper.

Let $\Phi(t)$ be the time-dependent solution of the wave equation
\begin{equation}
i\frac{\partial\Phi(t)}{\partial t}=H(t)\Phi(t)\,,
\end{equation}
our point of departure to obtaining $\Phi(t)$ is through the time-dependent
variational principle 
\begin{equation}
\left<\delta\Phi(t)\left\vert \left[i\frac{\partial}{\partial t}-H(t)\right]\right\vert \Phi(t)\right>=0\,.\label{eq:TDHF-1}
\end{equation}
Without loss of generality, it is convenient to factor out from $\Phi(t)$
the phase due to the time-evolution of the stationary state of $H_{0}$

\begin{equation}
\Phi(t)=e^{-iEt}\Psi(t)\,,
\end{equation}

\noindent with $E$ denoting the energy eigenvalue of $H_{0}$. As
a result, the time-dependent variational principle is recast as 
\begin{equation}
\left<\delta\Psi(t)\left\vert \left[E+i\frac{\partial}{\partial t}-H(t)\right]\right\vert \Psi(t)\right>=0\,.\label{eq:TDHF-2}
\end{equation}

For an\textit{ }atomic state with angular momentum \textit{JM} and
parity $\pi$, the multiconfiguration Hartree-Fock approximation assumes
the wave function $\Psi(t)$ as a superposition of configuration wave
functions $\psi_{a}(t)$ of the same \textit{JM} and $\pi$, viz.
\begin{equation}
\Psi(t)={\displaystyle \sum_{a}C_{a}(t)\psi_{a}(t)\,,}\label{eq:TDHF-2_wf}
\end{equation}
where \textit{a} is a configuration index, and $C_{a}(t)$ are time-dependent
weights. The configuration wave functions $\psi_{a}(t)$ are built
up from one-electron orbitals $u_{\alpha}(t)$. To guarantee the normalization
of $\Psi(t)$

\noindent 
\begin{equation}
\left<\Psi(t)\vert\Psi(t)\right>=1\,,\label{eq:wf_norm}
\end{equation}
the following subsidiary conditions 
\begin{align}
\left<u_{\alpha}(t)\vert u_{\beta}(t)\right> & =\delta_{\alpha\beta}\,,\\
\left<\psi_{a}(t)\vert\psi_{b}(t)\right> & =\delta_{ab}\,,\\
\sum_{a}C_{a}^{\star}(t)C_{a}(t) & =1\,,
\end{align}
are imposed. Since the perturbation $V(t)$ that induces atomic transitions
is harmonic in time, both $C_{a}(t)$ and $u_{\alpha}(t)$ assume
the following expansion

\begin{equation}
C_{a}(t)=C_{a}+\left[C_{a}\right]_{+}e^{-i\omega t}+\left[C_{a}\right]_{-}e^{+i\omega t}\ldots,\label{eq:RPA_C}
\end{equation}

\begin{equation}
u_{\alpha}(t)=u_{\alpha}+w_{\alpha+}e^{-i\omega t}+w_{\alpha-}e^{+i\omega t}+\ldots,\label{eq:RPA_u}
\end{equation}

\noindent where ``$\ldots$'' denotes higher harmonic responses.

Approximate time-dependent solution of Eq.~(\ref{eq:TDHF-2}) are
thus obtained with the wave function, Eq.~(\ref{eq:TDHF-2_wf}),
constrained by Eqs.~(\ref{eq:wf_norm}--\ref{eq:RPA_u}). The terms
$C_{a}$ and $u_{\alpha}$, which are independent of the external
field, lead to the usual stationary multiconfiguration Dirac-Fock
(MCDF) description of an atomic state and this gives the initial state
wave function $\ket{\Psi_{i}}$ of our problem. The terms $\left[C_{a}\right]_{\pm}$
and $w_{\alpha\pm}$, which are induced by the external field, lead
to equations describing the linear response of the atomic state to
the external field; these linear response equations are called the
multiconfiguration relativistic random-phase approximation (MCRRPA)
equations~\cite{Huang:1981wj,Johnson:1982hu}. Furthermore, the incoming
wave boundary condition is incorporated to yield the physical asymptotic
Coulomb wave function that describes an outgoing continuum electron
with a residual ion in the final state $\bra{\Psi_{f}}$ of our problem.
Since the external perturbation $V(t)$ may have components with nonvanishing
angular momentum and with odd parity, the atomic wave function contains
terms of mixed angular momentum and parity. If one starts from a single-configuration
reference state, the MCRRPA equations reduce to the usual RRPA equations.

To make connection with the general scattering formalism set up in
the last section, we note that the perturbing field components $v_{\pm}$
take the matrix form 
\begin{equation}
\left<\Psi_{f},\, E_{f}=E_{i}\pm\omega\left|v_{\pm}^{(w,\gamma)}\right|\Psi_{i}\right>=\left.\mathcal{M}^{(w,\gamma)}\right|_{E_{f}=E_{i}\pm\omega}\,,\label{eq:MCRRPA-scat.amp.}
\end{equation}
and they cause atomic excitation and de-excitation by an energy quantum
$\omega$, respectively. Therefore, in the process of solving the
MCRRPA equations, the corresponding scattering amplitudes are simultaneously
determined. 

Another important point to mention in our implementation of the MCRRPA
scheme is the choice of the spherical-wave basis. As a result, all
transition operators are cast into spherical multipole operators.
The key step in the spherical multipole expansion is breaking down
the atomic charge density operator $\hat{\mathcal{J}^{0}}(-\vec{q})$
into a series of charge multipole operators $\hat{C}_{JM}$ , and
the atomic 3-current density operator $\hat{\mathcal{J}}^{i}(-\vec{q})$
into a series of longitudinal $\hat{L}_{JM}$, transverse electric
$\hat{E}_{JM}$, and transverse magnetic $\hat{M}_{JM}$ multipole
operators, with $JM$ denoting the angular momentum quantum numbers.
The advantages of such an implementation include: (i) The three-dimensional
equation of motion for each orbital is reduced to a one-dimensional
equation. (ii) Each multipole operator has its own angular momentum
and parity selection rules, so the MCRRPA equations can be divided
into smaller blocks in which numerical calculations can be performed
more efficiently. (iii) For $1/|\vec{q}|$ larger than the size of
the atom, the multipole expansion converges rapidly. For the axial
charge $\hat{\mathcal{J}}_{5}^{0}(-\vec{q})$ and 3-current $\hat{\mathcal{J}}_{5}^{i}(-\vec{q})$,
four more types of multipole operators $\hat{C}_{JM}^{5}$, $\hat{L}_{JM}^{5}$,
$\hat{E}_{JM}^{5}$, and $\hat{M}_{JM}^{5}$, are to be introduced.
Each of them has opposite parity selection rule compared to its vector
counterpart. The details of the multipole expansion is given in Appendix~\ref{sec:app-1}.

\subsection{Atomic Structure of Germanium by MCDF \label{sub:Ge_MCDF}}

For the germanium atom, we chose the multiconfiguration reference
state to be 
\begin{equation}
\Psi=C_{1}\left(4p_{1/2}^{2}\right)_{0}+C_{2}\left(4p_{3/2}^{2}\right)_{0}\,,
\end{equation}
a linear combination of two configurations with total angular momentum
$J=0$ and parity $\pi=\mathrm{even}$, where the coefficients $C_{1}$
and $C_{2}$ are the configuration weights. The notation $\left(4l_{j}^{2}\right)$
denotes symbolically an anti-symmetrized wave function constructed
from two electrons in the valence orbital $4l_{j}$. The rest of the
electrons in the 10 inner orbitals $4s_{1/2}$, $3d_{5/2}$, $3d_{3/2}$,
$3p_{3/2}$, $3p_{1/2}$, $3s_{1/2}$, $2p_{3/2}$, $2p_{1/2}$, ,
$2s_{1/2}$, and $1s_{1/2}$ form the closed core.

The ground-state wave function obeying the MCDF equations is solved
by the computer code~\cite{Desclaux:1974jp}, which yields all the
core and valence orbitals, and the configuration weights $C_{1}$
and $C_{2}$. In Table~\ref{tab:Ge_s.p.}, all calculated orbital
binding energies are shown and compared with the edge energies extracted
from photoabsorption data of germanium solids (to be discussed in
the next section). In Table~\ref{tab:Ge_s.p.}, the configuration
weights and their corresponding percentages are given.

\begin{table}
\protect\caption{The binding energies (in eV) of the atomic germanium orbits from the
present MCDF calculations. The experimental data are the edge energies
extracted from photoabsorption data of germanium solids. \label{tab:Ge_s.p.}}

\medskip{}

\begin{tabular}{>{\centering}p{1.8cm}>{\centering}p{0.1cm}>{\centering}p{1.8cm}>{\centering}p{0.1cm}>{\raggedleft}m{1.8cm}>{\centering}p{0.1cm}>{\raggedleft}m{1.8cm}}
\hline 
\multicolumn{3}{c}{Label} &  & MCDF  &  & Exp.$^{a}$ \tabularnewline
\hline 
Subshell  &  & Orbital  &  &  &  & \tabularnewline
\hline 
$N_{III}$  &  & $4p_{1/2}$  &  & 7.8  &  & \tabularnewline
$N_{II}$  &  & $4p_{3/2}$  &  & 8.0  &  & \tabularnewline
$N_{I}$  &  & $4s_{1/2}$  &  & 15.4  &  & \tabularnewline
$M_{V}$  &  & $3d_{5/2}$  &  & 43.1  &  & 29.3 \tabularnewline
$M_{IV}$  &  & $3d_{3/2}$  &  & 43.8  &  & 29.9 \tabularnewline
$M_{III}$  &  & $3p_{3/2}$  &  & 140.1  &  & 120.8 \tabularnewline
$M_{II}$  &  & $3p_{1/2}$  &  & 144.8  &  & 124.9 \tabularnewline
$M_{I}$  &  & $3s_{1/2}$  &  & 201.5  &  & 180.1 \tabularnewline
$L_{III}$  &  & $3p_{3/2}$  &  & 1255.6  &  & 1217.0 \tabularnewline
$L_{II}$  &  & $3p_{1/2}$  &  & 1287.9  &  & 1248.1 \tabularnewline
$L_{I}$  &  & $2s_{1/2}$  &  & 1454.4  &  & 1414.6 \tabularnewline
$K$  &  & $1s_{1/2}$  &  & 11185.5  &  & 11103.1 \tabularnewline
\hline 
\multicolumn{7}{l}{$^{a}$From Ref.~\cite{Henke:1993gd}.}\tabularnewline
\end{tabular}
\end{table}

\begin{table}
\protect\caption{Configuration weights of the germanium atom in its ground state ($J^{\pi}=0^{+}$)
from the present MCDF calculations. \label{tab:Ge_m.c.}}

\medskip{}
\begin{tabular}{>{\centering}p{4cm}>{\centering}p{0.1cm}>{\centering}p{4cm}cc}
\hline 
Valence Configuration  &  & Configuration Weight  &  & Percentage\tabularnewline
\hline 
$4p_{1/2}^{\,2}$  &  & 0.84939  &  & 72.15$\%$ \tabularnewline
$4p_{3/2}^{\,2}$  &  & 0.52776  &  & 27.85 $\%$\tabularnewline
\hline 
\end{tabular}
\end{table}

\subsection{Photoabsorption of Germanium by MCRRPA \label{sub:Ge_photo}}

To further benchmark the MCRRPA method, in particular its applicability
to the atomic bound-to-free transition of germanium, we consider the
photoabsorption of germanium above the ionization threshold, for which
experimental data are available.

In the multipole expansion scheme, an external perturbing field with
parameters $J$ and $\lambda$ gives rise to one-particle-one-hole
excitation channels which are restricted by the angular momentum and
parity conservation. Suppose one of the atomic bound electron in the
$nl_{j}$ orbital is promoted to a free continuum state $\epsilon l_{j^{\prime}}^{\prime}$
($\epsilon$ denotes the kinetic energy) by this $J\lambda$ perturbing
field, the relevant quantum numbers then satisfy the following selection
rules:

\begin{equation}
\left|\, j-J\right|\,\leqslant\, j^{\prime}\,\leqslant\left|\, j+J\right|,\quad(\mathrm{Angular\: Momentum\: Selection\: Rule})
\end{equation}

\begin{equation}
l+l^{\prime}+J+\lambda-1=\mathrm{even}.\quad(\mathrm{Parity\: Selection\: Rule})
\end{equation}
As a result, in response to the multipole perturbations (with different
$J\lambda$), the germanium atom (a many-body $^{1}S_{0}$ state)
is excited to a state mixed with components of different total angular
momenta and parities.

For example, consider the case arising from excitations of the two
valence electrons in the valence orbitals \textit{$4p_{1/2}$} or
\textit{$4p_{3/2}$}. There are 5 possible excitation channels responding
to the electric-type dipole excitation (by a $E_{J=1}$ operator):

\begin{center}
$\qquad\qquad4p_{1/2}\rightarrow\epsilon s_{1/2},$ 
\par\end{center}

\begin{center}
$\qquad\qquad4p_{1/2}\rightarrow\epsilon d_{3/2},$ 
\par\end{center}

\begin{center}
$\qquad\qquad4p_{3/2}\rightarrow\epsilon s_{1/2},$ 
\par\end{center}

\begin{center}
$\qquad\qquad4p_{3/2}\rightarrow\epsilon d_{3/2},$ 
\par\end{center}

\begin{center}
$\qquad\qquad4p_{3/2}\rightarrow\epsilon d_{5/2}.$ 
\par\end{center}

Besides the above valence-excitation channels, the 10 inner core orbitals
give rise to additional 24 channels. In total, when one considers
all possible excitations from all orbitals, there are 29 excitation
channels to be taken into account in the electric-type dipole excitation.
These 29 interacting \textit{jj}-coupled channels are all included
in our MCRRPA framework to account for the final ionic-state electron
correlations. The corresponding MCRRPA equations comprise a system
of coupled differential equations up to 29 channels with 116 unknown
radial functions to be numerically solved in a self-consistent manner.

To obtain the total photoabsorption cross section, all electric-type
($E_{J}$) and magnetic-type ($M_{J}$) multipole excitations which
contribute to the on-shell transverse response function, $R_{11+22}^{(\gamma)}(q^{2}=0)$,
are summed. For photons with energy $T\lesssim10\,\mathrm{keV}$,
it is found that high-order multipole transition probabilities decrease
rapidly in an exponential\textbf{ }mode. We choose the cut-off value
$J_{\mathrm{cut}}$ in the multipole expansion by the following recursive
procedure: We first sum over the multipole transition probabilities
up to a definite polarity order (which should be high enough so the
rapidly-decreasing pattern starts to show), and extrapolate the corrections
from succeeding higher multipoles by a proper exponential form. Then
$J_{\mathrm{cut}}$ is fixed once the contributions from $\sum_{J>J_{\mathrm{cut}}}$
is estimated, by the exponential law, to be below 1\% of the total
from $\sum_{J\leq J_{\mathrm{cut}}}$.

\begin{figure}[h]
\includegraphics[width=0.49\textwidth]{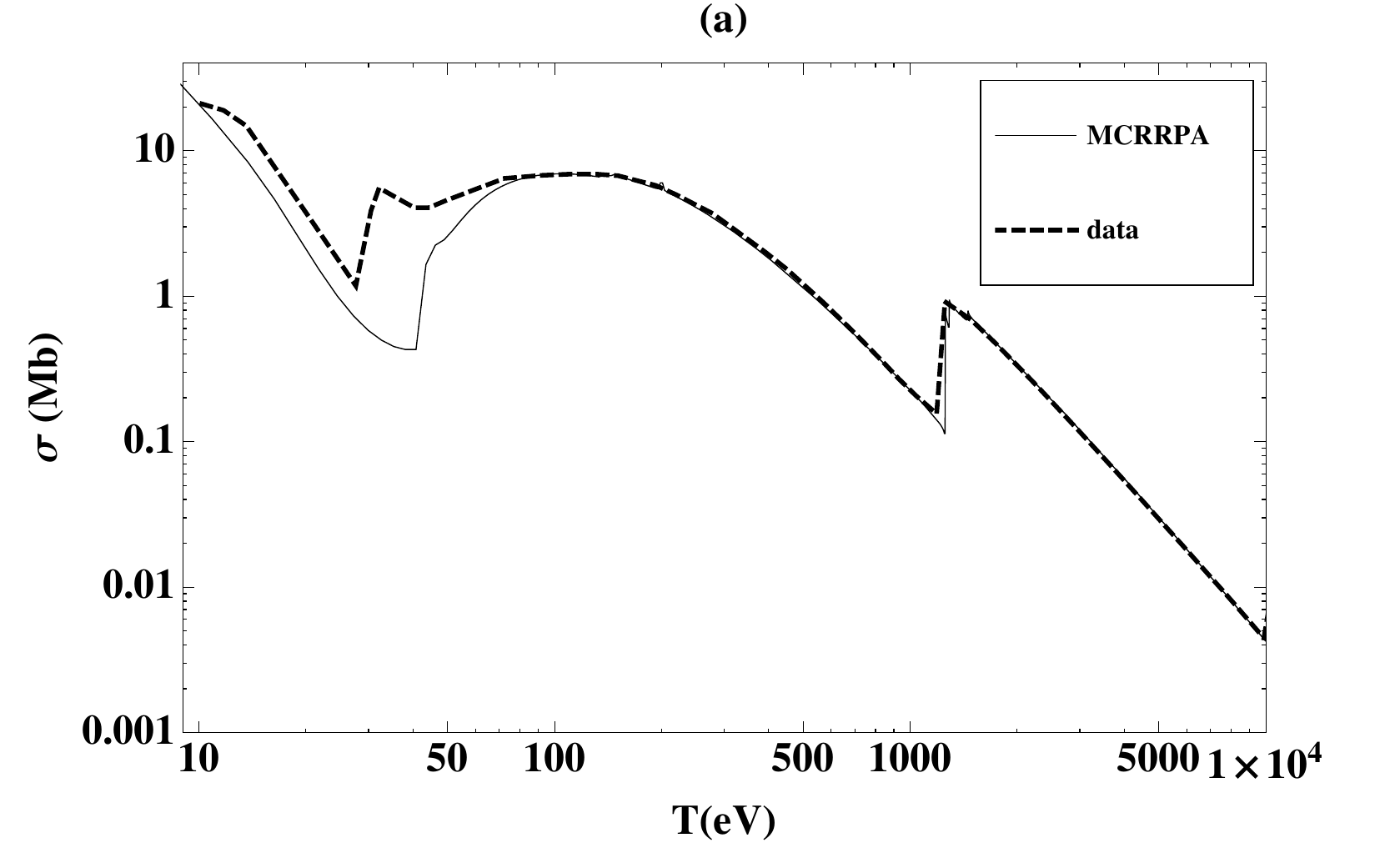} \includegraphics[width=0.49\textwidth]{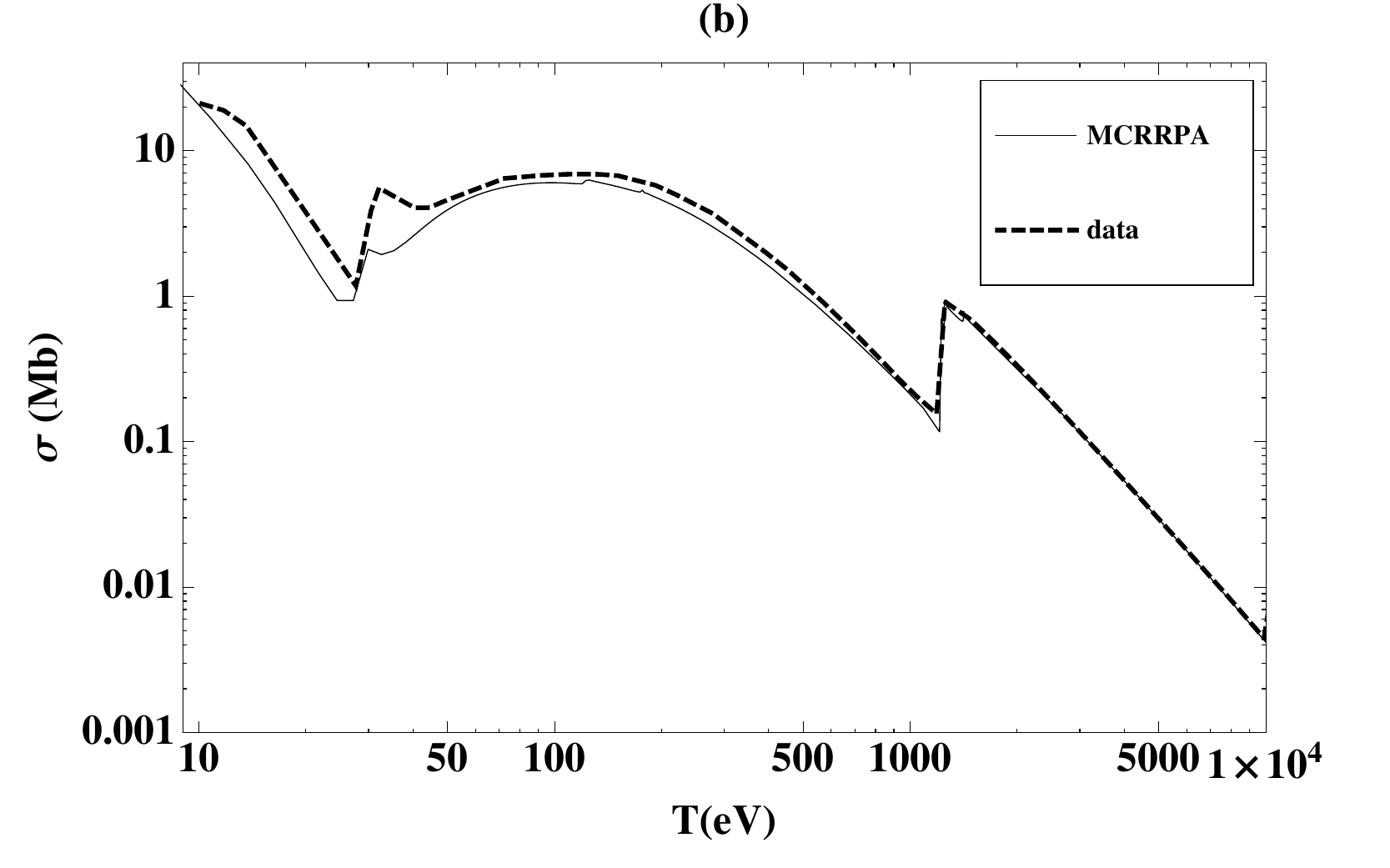}
\protect\caption{Photoabsorption cross sections of Ge. The data are taken from Ref.~\cite{Henke:1993gd}.
The MCRRPA line in panel (a) shows our numerical results; the one
in panel (b) is obtained by forcing all shell energies aligned to
the experimental edge energies. \label{fig:sig_gamma_Ge}}
\end{figure}

In Fig.~\ref{fig:sig_gamma_Ge}(a), the photoabsorption cross sections
from the MCRRPA method and experimental data are shown for incident
photon energies ranging from 10 eV to 10 keV. The MCRRPA results agree
very well with experiments for photon energies larger than 80 eV,
with errors uniformly below the $5\%$ level. The discrepancy below
80 eV is relatively large and we believe it is due to the fact that
the experimental data were taken from solid-phase Ge targets, whose
wave functions and orbital binding energies, in particular for outer-shell
electrons, are affected by nearby atoms and therefore different from
the ones of a single atom. As shown by Table~\ref{tab:Ge_s.p.} and
Fig.~\ref{fig:sig_gamma_Ge}(a), the solid effects are especially
significant for the 3\textit{d} orbitals. On the other hand, the inner-shell
electrons are less affected by crystal structure; as a result, our
calculation well-reproduces the data of photon energies $T\ge100\,\mathrm{eV}$,
where cross sections are dominated by ionization of inner-shell electrons.
To estimate the degree to which our MCRRPA results will be affected
by the solid effects in the $T\ge100\,\mathrm{eV}$ region, we carried
out a parallel calculation in which the theoretical ionization thresholds
are artificially aligned with the experimental ones. The results,
plotted in Fig.~\ref{fig:sig_gamma_Ge}(b), show that the deviations
from experimental data are still kept below the $10\%$ level. Therefore,
we estimate the theoretical uncertainty due to the solid effects to
be $\lesssim10\%$ in the $T\ge100\,\mathrm{eV}$ region.

Summing up this section, we demonstrate that our MCRRPA approach is
capable of giving a good description of a germanium atom and its photoabsorption
process with photon energy larger than 100 eV. In other words, the
many-body wave functions, single particle basis states, and relevant
transition matrix elements thus obtained should be good approximations
to the exact answers. In the next section, we shall apply this approach
to germanium ionization by neutrinos.

\section{Ionization of Germanium by Neutrinos}

As shown in Eqs.~(\ref{eq:dS/dT_weak},\ref{eq:dS/dT_F1},\ref{eq:dS/dT_F2}),
ionization of germanium by neutrinos depends on various atomic response
functions $R$'s, which need explicit many-body calculations. The
only differences in calculating the response functions for this case
from the ones for photoionization are (i) different atomic current
operators are involved, and (ii) different kinematics are probed (the
former are mostly off-shell, while the latter are purely on-shell).
Therefore, it is straightforward to treat the problem in the MCRRPA
framework simply by taking more types of multipole operators and their
off-shellness into account. Both aspects are not expected to generate
additional complexity or problems in many-body physics, therefore,
one can take similar confidence on MCRRPA in this case as what has
been acquired in the photoabsorption case with $T\ge100\,\mathrm{eV}$.

\begin{figure}[h]
\noindent \begin{centering}
\includegraphics[width=0.32\textwidth]{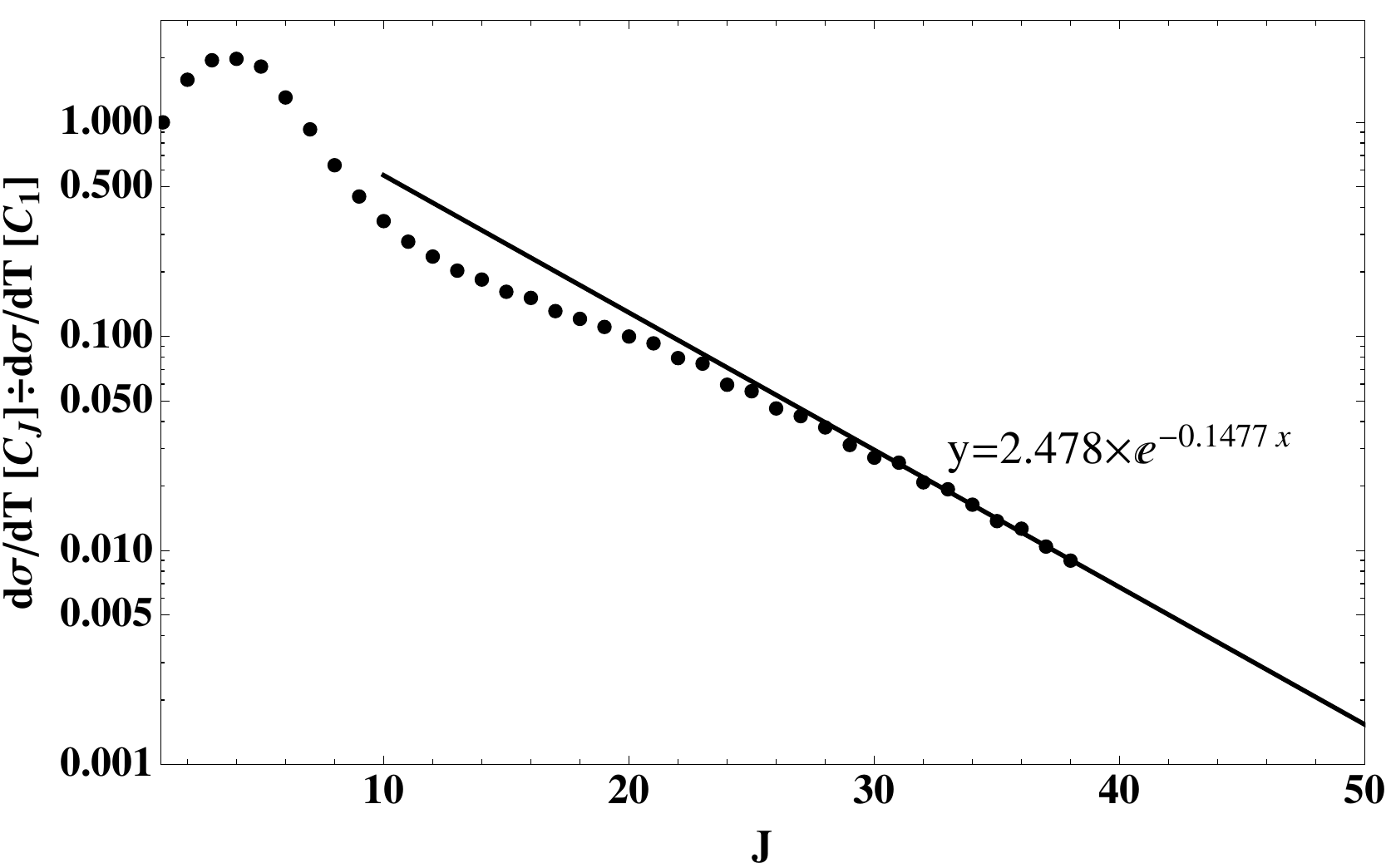}\includegraphics[width=0.32\textwidth]{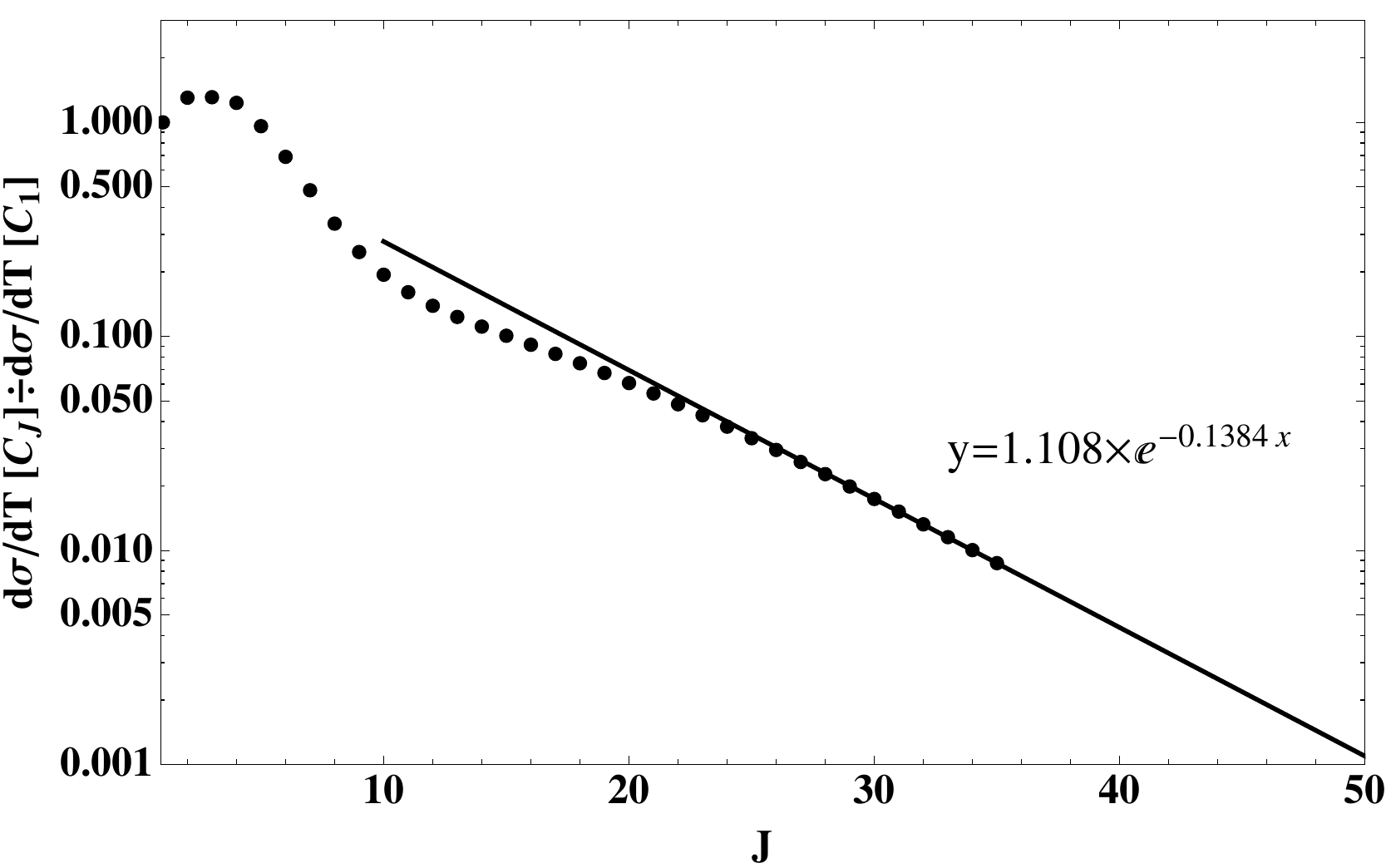}\includegraphics[width=0.32\textwidth]{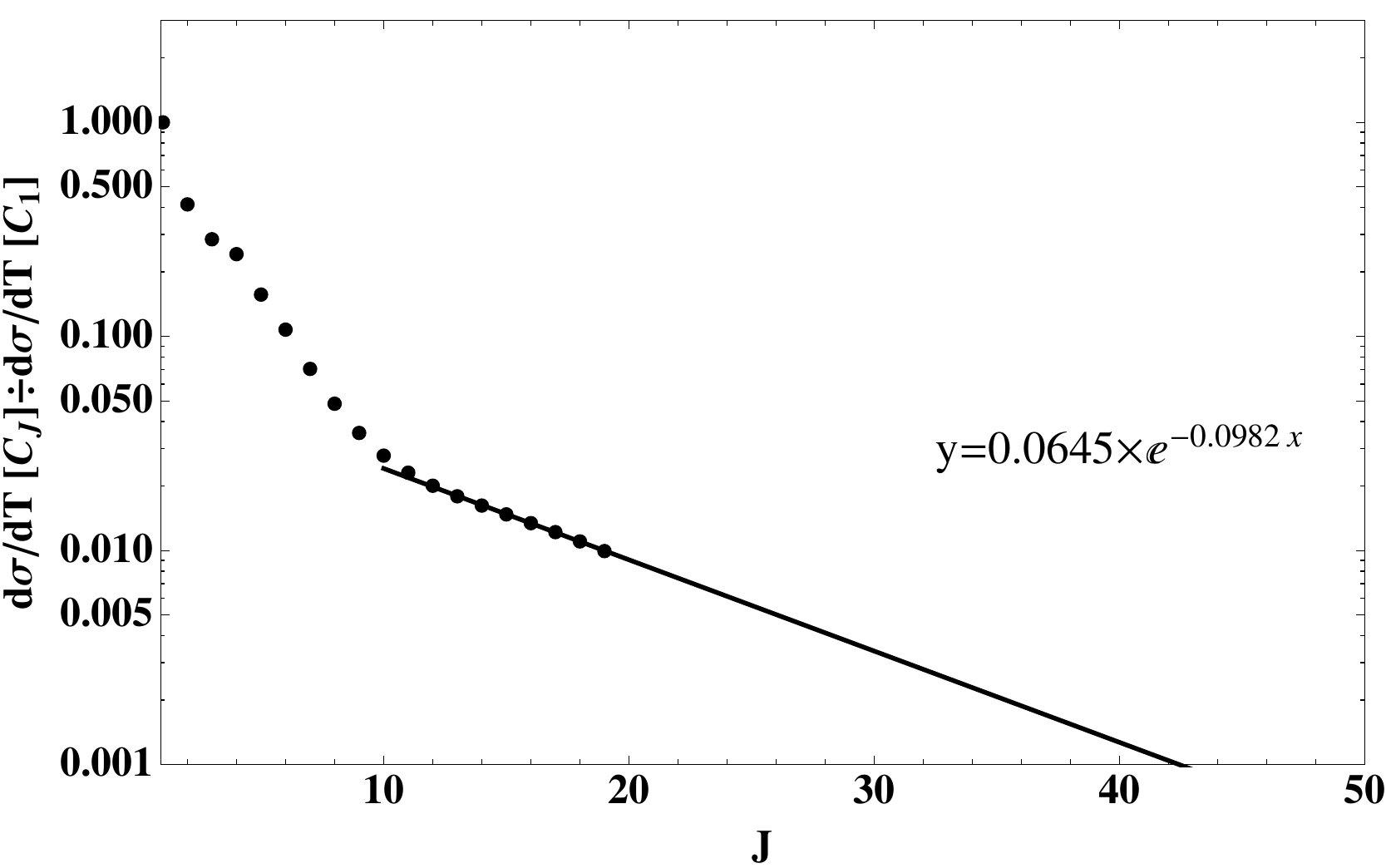} 
\par\end{centering}

\noindent \protect\caption{Normalized contributions from the series of charge multipole operators
$C_{J}$'s to differential cross sections for (a) weak interaction,
(b) magnetic moment interaction, and (c) millicharge interaction.
The incident neutrino has 1 MeV energy and deposits 1 keV energy.
\label{fig:mul_conv}}
\end{figure}

Because $q^{2}$ in a $t$-channel scattering process is space-like,
i.e., $q^{2}<0$ or $T^{2}<|\vec{q}|^{2}$, an off-shell current operator
typically yields a multipole expansion which converges more slowly
than its on-shell counterpart. Here we use an example to demonstrate
a multipole expansion scheme is still valid and effective for the
cases we are interested. Consider an incident neutrino with 1 MeV
energy (a typical value for reactor antineutrinos) and depositing
1 keV energy to the detector through the charge-type multipole operators
$\hat{C}_{JM}$ in weak, magnetic moment, or millicharge interactions.
The contributions of $\hat{C}_{JM}$ to the differential cross sections
$d\sigma/dT$ in these three cases are plotted in Fig.~\ref{fig:mul_conv}(a),
(b), and (c), respectively. All these plots feature exponential decay
behaviors with increasing multipolarity $J$, and they are fitted
to be proportional to $e^{-0.15J}$, $e^{-0.14J}$, and $e^{-0.10J}$,
respectively. Therefore, we can apply the same cutoff procedure mentioned
in the last section in multipole expansions and control the higher-multipole
uncertainty at the $1\%$ level. For the entire kinematics considered
in this work, it is found that the cutoff values $J_{\mathrm{cut}}$
are no more than $50-60$.

\subsection{Results and Discussion \label{sub:Ge_nuAI}}

In this section, we present our calculated differential cross sections
for germanium ionization with two representative incident neutrino
energies: (a) $E_{\nu}=1\,\mathrm{MeV}$ and (b) $E_{\nu}=10\,\mathrm{keV}$.
The former case is typical for reactor antineutrinos, while the latter
case gives an example of very low energy neutrinos, e.g., ones from
tritium $\beta$ decay.

\subsubsection{Weak Interaction}

\begin{figure}[h]
\includegraphics[width=0.49\textwidth]{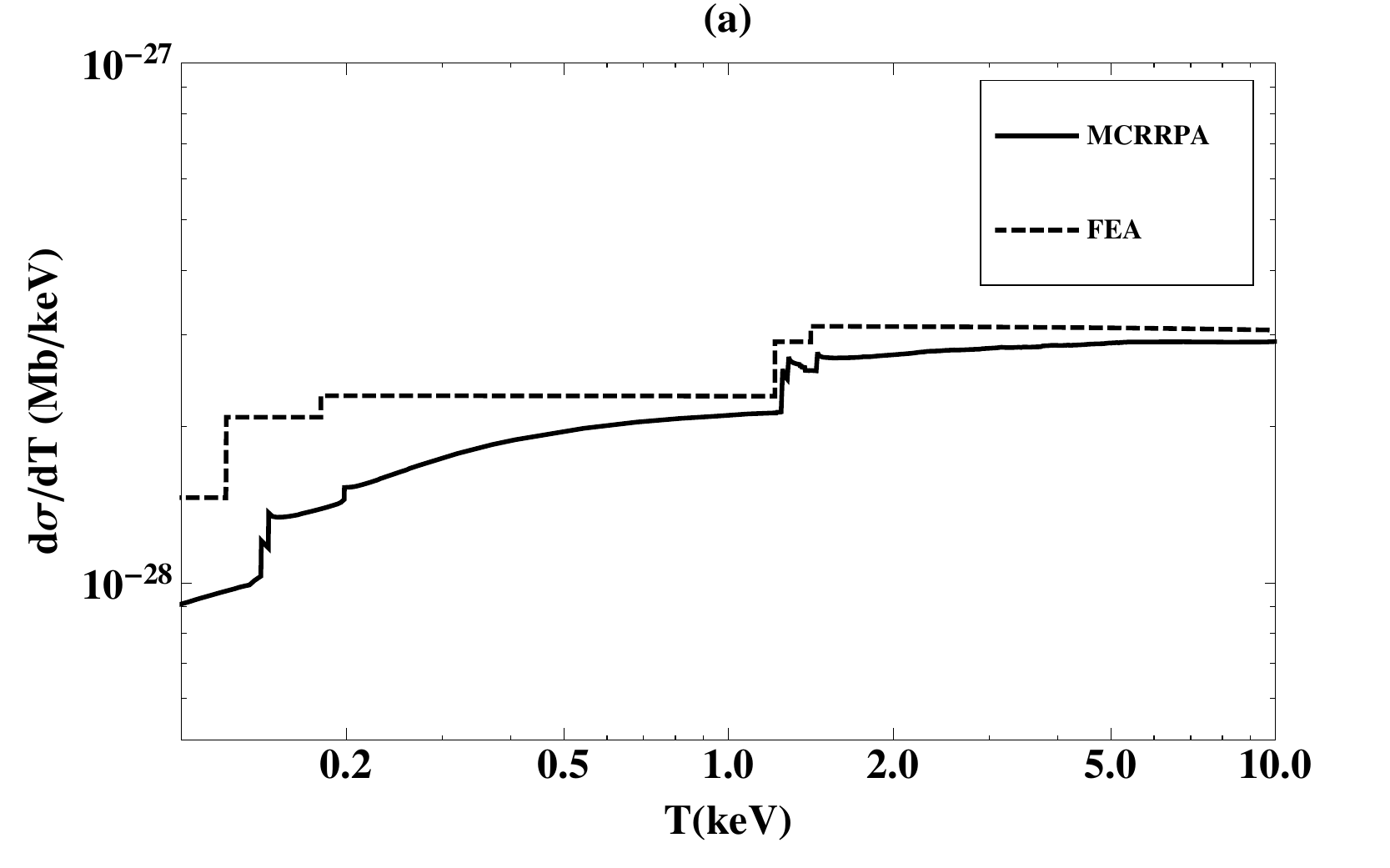}\includegraphics[width=0.49\textwidth]{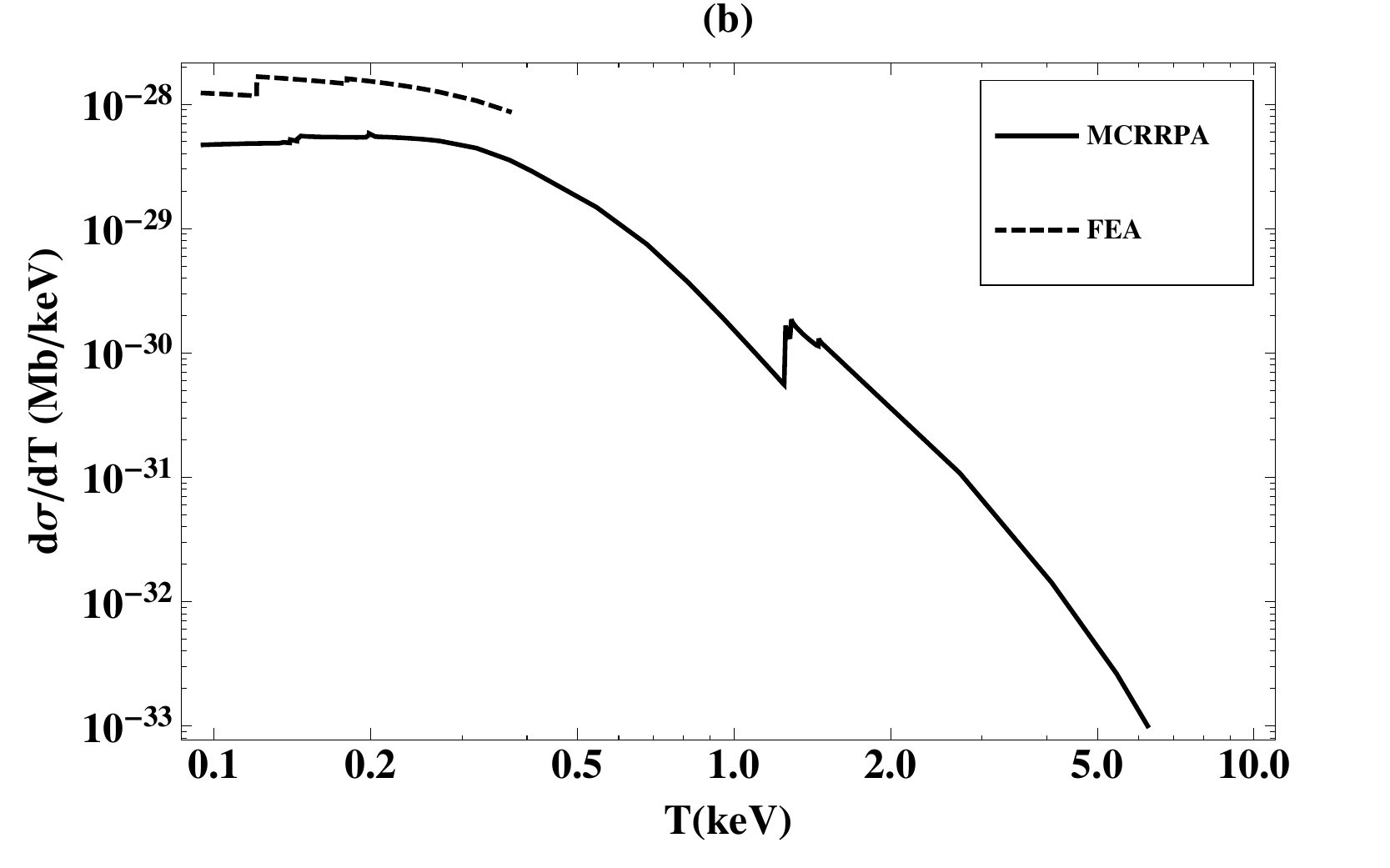}
\protect\caption{Differential cross sections for germanium ionization by neutrino weak
interaction with neutrino incident energies (a) $E_{\nu}=$1 MeV and
(b) $E_{\nu}=$ 10 keV. (See also Fig.~2 in Ref.~\cite{Chen:2013lba})
\label{fig:ds/dT_weak}}
\end{figure}

The differential cross sections due to the weak interaction, i.e.,
Eq.~(\ref{eq:dS/dT_weak}), are given in Fig.~\ref{fig:ds/dT_weak}
(see also Fig.~2 in Ref.~\cite{Chen:2013lba}). As shown in panel
(a), our MCRRPA calculation and the conventional FEA scheme gradually
converge when the energy transfer is larger than 1 keV. On the other
hand, below $T=1\,\mathrm{keV},$ FEA starts to overestimate the differential
cross sections. In other words, we found the atomic binding effect
suppress the weak scattering cross sections at low energies in comparison
to the free scattering picture. This conclusion is consistent with
previous explicit many-body calculations~\cite{Fayans:1992kk,Kopeikin:1997ge,Fayans:2001pg,Kopeikin:2003bx}.

In very low energy neutrino scattering, the FEA scheme has another
severe problem that comes with its specific kinematic constrain: $q^{2}=-2m_{e}T$.
This leads to a maximum energy transfer $T_{\mathrm{max}}\approx0.38\,\mathrm{keV}$
for a 10-keV neutrino beam--as shown by the sharp cutoff for the FEA
curve in panel (b); while there is no such cutoff expected in a neutrino--atom
ionization process. Experiments with good energy resolution should
be able to discern this difference.

\subsubsection{Magnetic Moment Interaction}

\begin{figure}[h]
\includegraphics[width=0.49\textwidth]{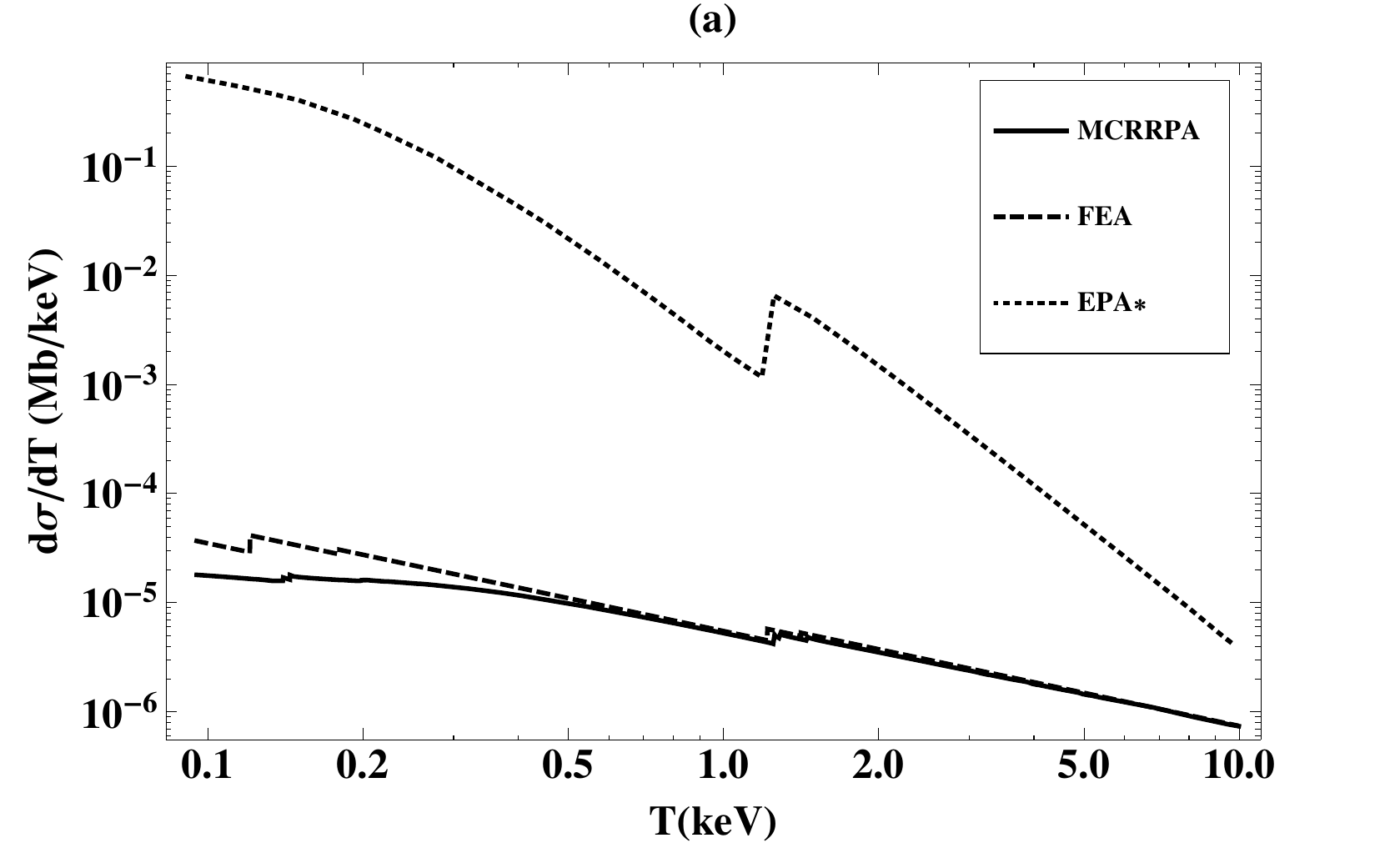}\includegraphics[width=0.49\textwidth]{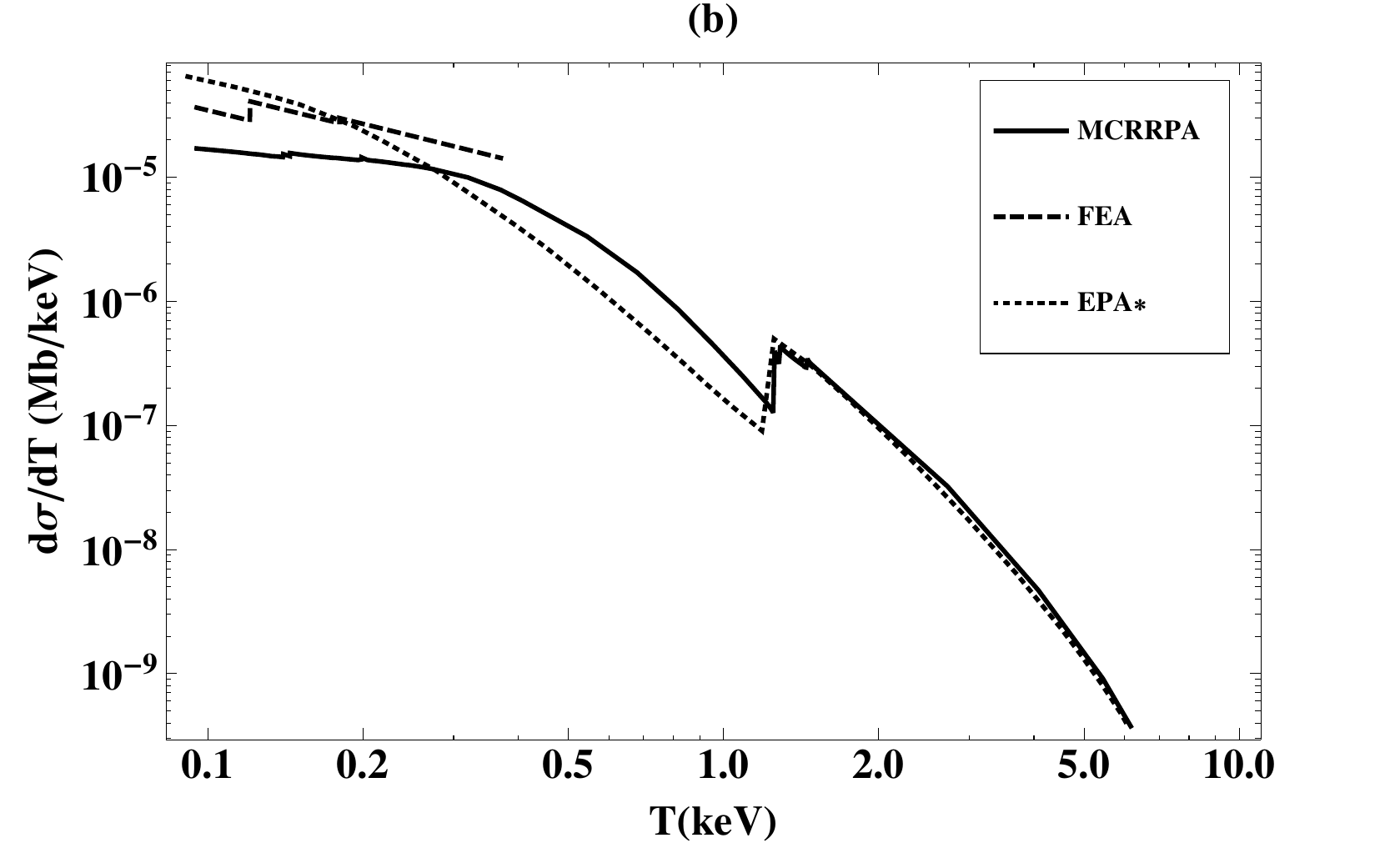}
\protect\caption{Differential cross sections for germanium ionization by neutrino magnetic
moment interaction with neutrino incident energies (a) $E_{\nu}=$1
MeV and (b) $E_{\nu}=$ 10 keV, in units of \textbf{$\bbkappa_{\nu}^{\mathrm{(eff)}2}$}.
(See also Fig.~2 in Ref.~\cite{Chen:2013lba}) \label{fig:ds/dT_NMM}}
\end{figure}

The differential cross sections due to the interaction with $\bbkappa_{\nu}^{(\mathrm{eff})}$,
i.e., Eq.~(\ref{eq:dS/dT_F2}), are given in Fig.~\ref{fig:ds/dT_NMM}
(see also Fig.~2 in Ref.~\cite{Chen:2013lba}). The comparison of
the MCRRPA and FEA results shows very similar features as the case
of weak scattering: FEA overestimates in the $T\lesssim1\,\mathrm{keV}$
region and gradually converges to MCRRPA for $T\gtrsim1\,\mathrm{keV}$,
and our conclusion in this case is also consistent with previous explicit
many-body calculations~\cite{Fayans:1992kk,Kopeikin:1997ge,Fayans:2001pg,Kopeikin:2003bx}.

As there have been quite extensive recent discussions about the role
of atomic structure in scattering by neutrino magnetic moments, we
try to clarify the confusion which is caused by the applicabilities
of various approximation schemes:
\begin{enumerate}
\item EPA: It was first claimed in Ref.~\cite{Wong:2010pb} that atomic
structure can greatly enhance the sensitivity to $\bbkappa_{\nu}^{(\mathrm{eff})}$
by orders of magnitude than the FEA prediction in the $T<1\,\mathrm{keV}$
region for germanium. However, later works inspired by this, using
various approaches, all came to the opposite conclusion~\cite{Voloshin:2010vm,Kouzakov:2010tx,Kouzakov:2011vx,Chen:2013lba}.
The source of the huge overestimation in Ref.~\cite{Wong:2010pb}:
the use of an unconventional EPA scheme, was pointed out in Ref.~\cite{Chen:2013iud}
by considering a simple case of hydrogen atoms. Applying the same
scheme to germanium, the results are shown by the EPA$^{*}$ curves
in Fig.~\ref{fig:ds/dT_NMM}. In panel (a), one clearly sees the
orders-of-magnitude enhancement that EPA$^{*}$ predicts. On the other
hand, in panel (b), EPA$^{*}$ does agree well with MCRRPA for $T>1\,\mathrm{keV}$.
This is consistent with the feature pointed out in Ref.~\cite{Chen:2013iud}:
When incident neutrino energy (in this case, 10 keV) falls below the
scale of atomic binding momentum (in this case, 35 keV for the most
important $3p$ shell), the EPA$^{*}$ works incidentally.
\item The Voloshin sum rules: Quantum-mechanical sum rules for neutrino
weak and magnetic-moment scattering were derived by Voloshin~\cite{Voloshin:2010vm}
and refined in later works~\cite{Kouzakov:2010tx,Kouzakov:2011vx}.
Using several justified assumptions, the sum rules concluded that
treating atomic electrons as free particles be a good approximation.
One important step in these sum rules is extending the integration
over $q^{2}$ (equivalent to integration over the neutrino scattering
angle $\theta$ for a fixed $T$) from the physical range $[T^{2},4E_{\nu}^{2}]$
to $[0,\infty)$. In this sense, the sum-rule results, or equivalently
the FEA results, can be interpreted as upper limits for realistic
$d\sigma/dT$, and this is consistent with our MCRRPA curves being
under the FEA ones in Figs.~\ref{fig:ds/dT_weak} and \ref{fig:ds/dT_NMM}.
However, the larger discrepancy between realistic calculations and
FEA at sub-keV energies seems to be in contradiction with the sum-rule-FEA
argument: With low $T$, only outer-shell electrons are ionized, so
the sum rules should work even better, not worse, since these electrons
are less bound, or closer to be free electrons. The main reason, as
pointed out in Ref.~\cite{Kouzakov:2014lka}, is the missing of two-electron
correlation in the sum rule derivation, which plays a more important
role at low energies. 
\end{enumerate}

\subsubsection{Millicharge Interaction}

\begin{figure}[h]
\includegraphics[width=0.49\textwidth]{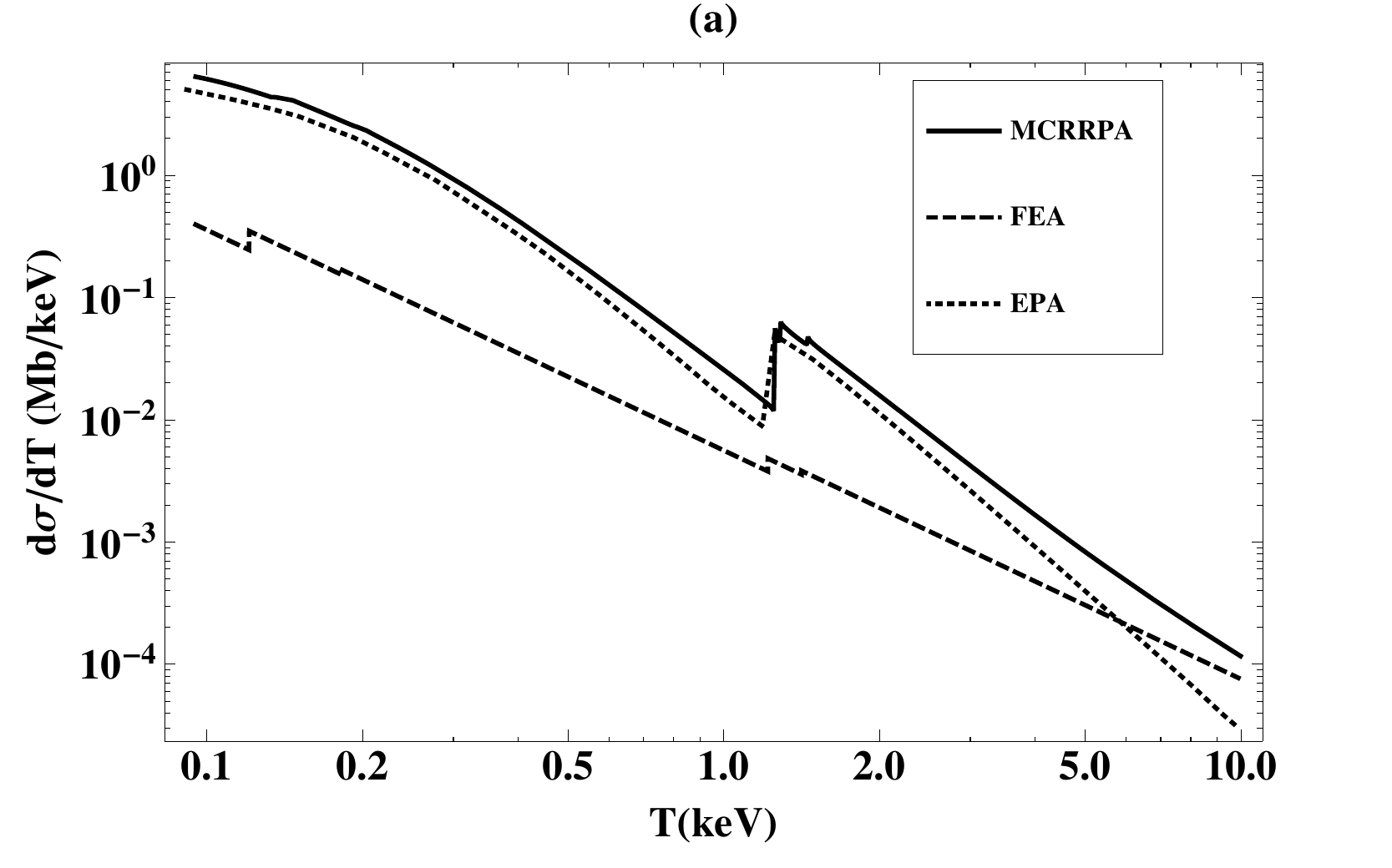}\includegraphics[width=0.49\textwidth]{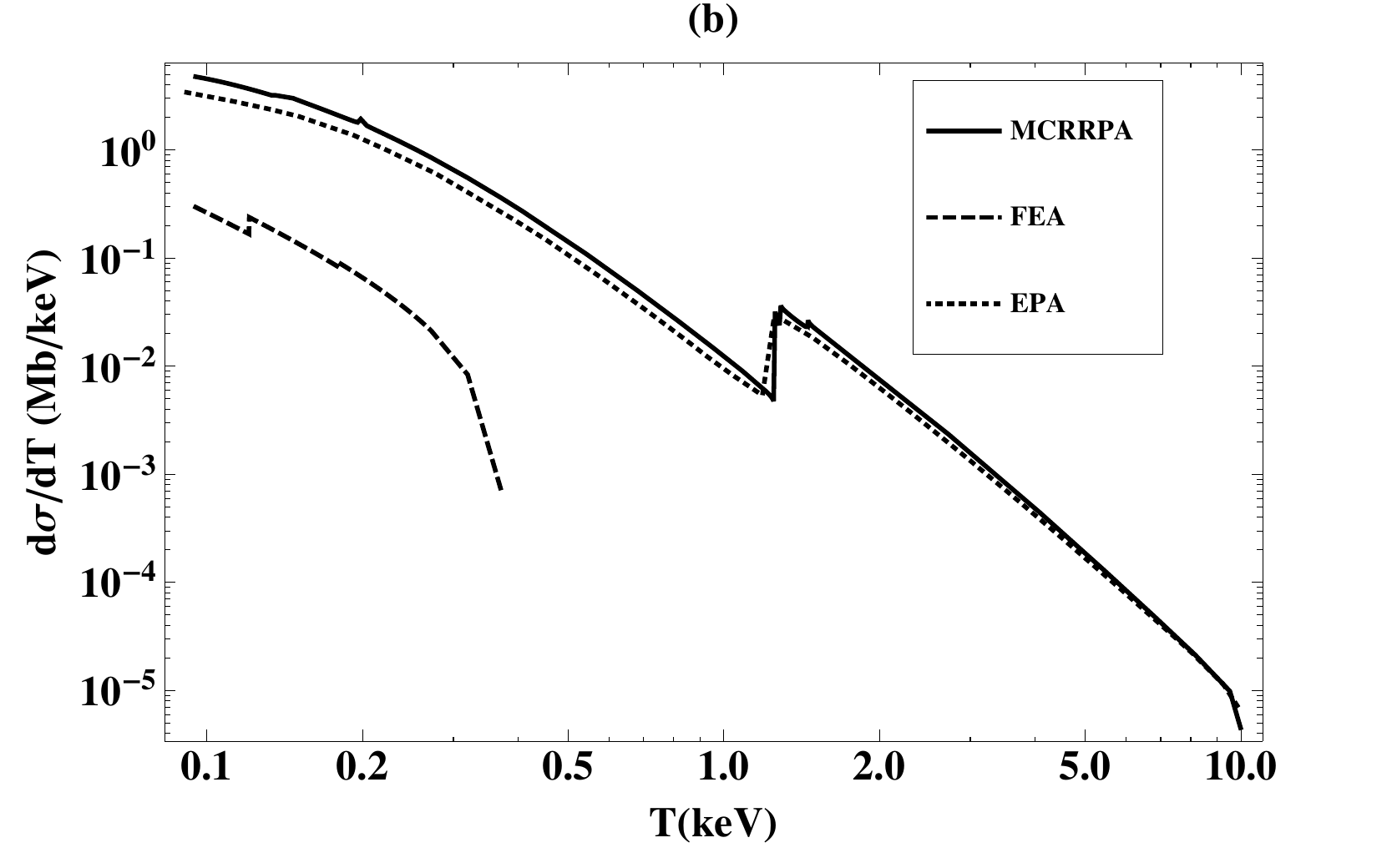}
\protect\caption{Differential cross sections for germanium ionization by neutrino millicharge
interaction with neutrino incident energies (a) $E_{\nu}=$1 MeV and
(b) $E_{\nu}=$ 10 keV, in units of $\mathbbm{q}_{\nu}^{2}$. \label{fig:ds/dT_mQ} }
\end{figure}

The differential cross sections due to the interaction quadratic in
\textbf{$\mathbbm{q}_{\nu}$}, i.e., Eq.~(\ref{eq:dS/dT_F1}), are
given in Fig.~\ref{fig:ds/dT_mQ}. While the linear term due to the
EM--weak interference can be calculated straightforwardly, it can
be safely ignored at the current and projected sensitivity levels
of direct experiments with \textbf{$\mathbbm{q}_{\nu}\sim10^{-12}-10^{-13}$}. 

Unlike the previous two cases that FEA works well for neutrino weak
and magnetic moment scattering with big enough incident energy $E_{\nu}$
and energy deposition $T$, it underestimates the millicharge scattering
cross sections, in particular in the most interested sub-keV region
of $T$. Instead, it is EPA that works much better in this case. The
main reason, as pointed out in Ref.~\cite{Chen:2013iud}, is due
to the kinematic factor $1/q^{2}$ that goes along the transverse
response function $R_{11+22}^{(\gamma)}$ in Eq.~(\ref{eq:dS/dT_F1}).
This factor weights more the forward scattering region with $q^{2}\rightarrow0$,
where photons behave like real particles. For the same reason one
can see that the FEA constraint: $q^{2}=-2m_{e}T$ deviates substantially
from the true kinematics of this scattering process. 

Because of the same $1/q^{2}$ factor, we also note that the differential
cross section contains a logarithmic term $\log(E_{\nu}/m_{\nu})$,
which diverges at the limit of massless neutrinos~\cite{Chen:2014dsa}.
While it is known that neutrinos are not massless, their masses have
not been determined precisely yet. Instead of using the current upper
limit $m_{\nu_{e}}<2\,\mathrm{eV}$ as the cutoff value in this logarithm
to present our results in this paper, we adopt the Debye length of
germanium solid: 0.68 $\mu\mathrm{m}$ which characterizes the scale
of screen Coulomb interaction and acts like a 0.29 eV mass cutoff
(a value also similar to the projected sensitivity on $m_{\nu_{e}}$
by the KATRIN experiment). The uncertainty in cross sections due to
this one-order-of-magnitude difference in the mass cutoff is about
$20\%$.

\subsubsection{Charge Radius Interaction}

\begin{figure}[h]
\includegraphics[width=0.49\textwidth]{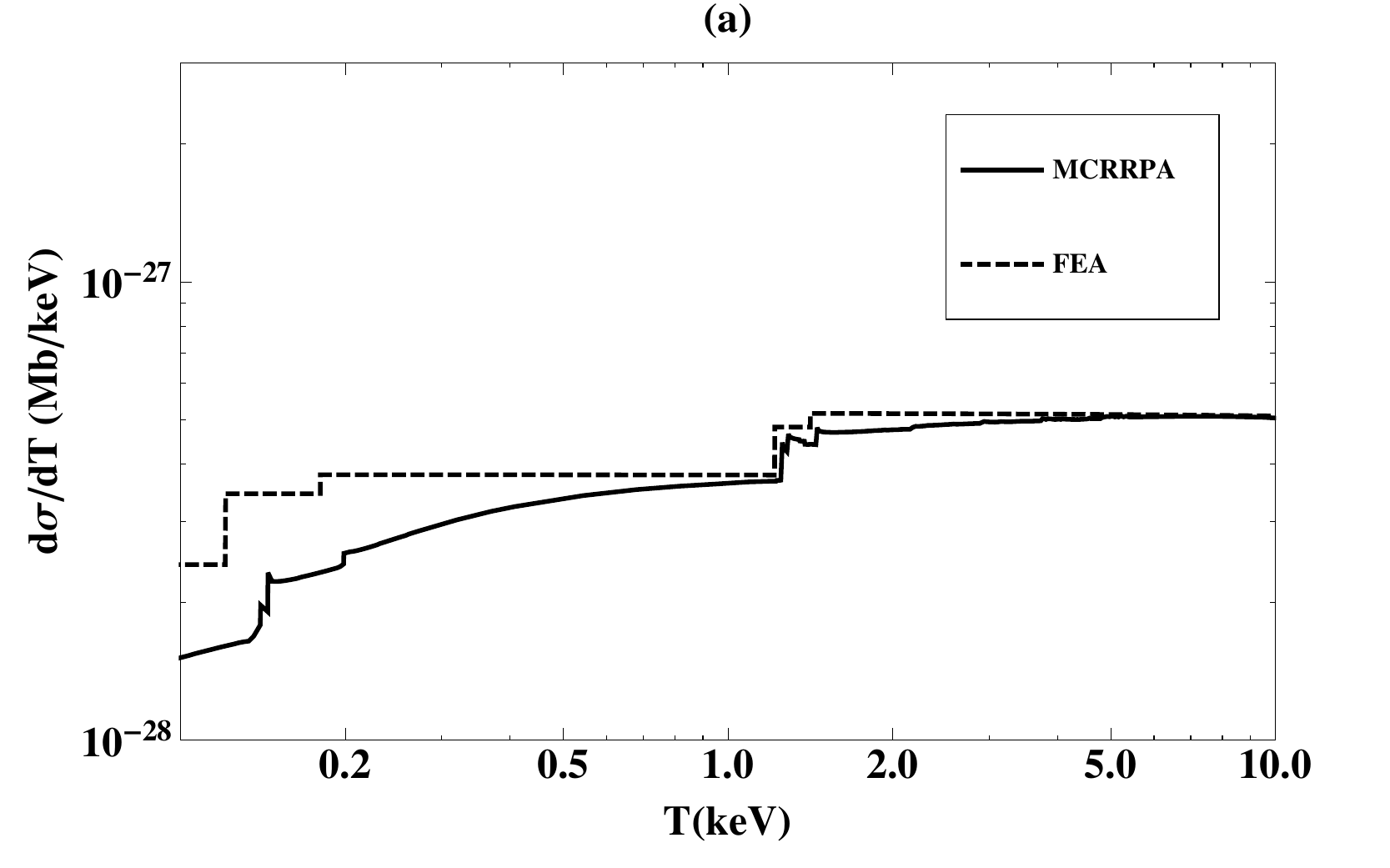}\includegraphics[width=0.49\textwidth]{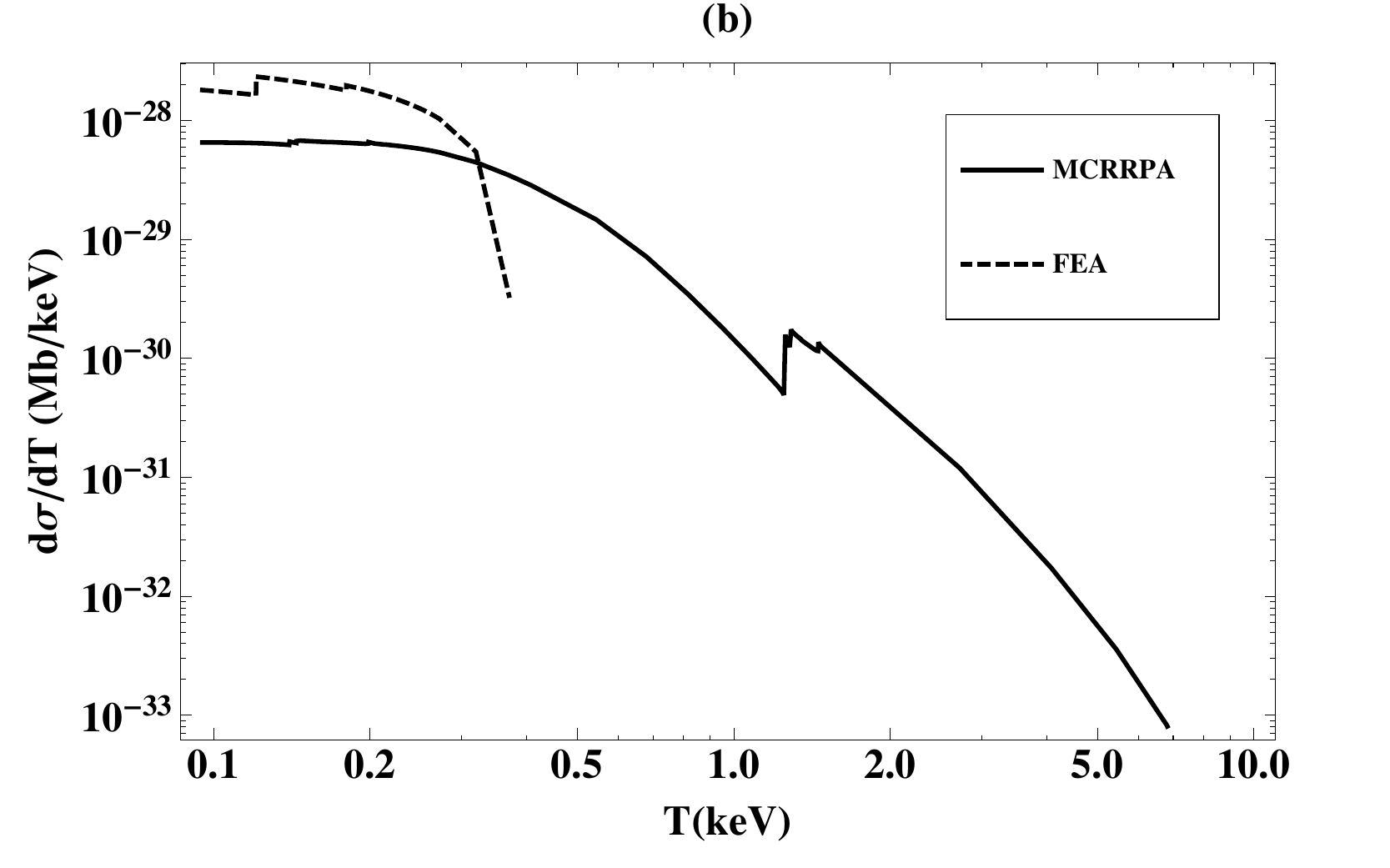}
\protect\caption{Differential cross sections for germanium ionization by neutrino charge
radius interaction with neutrino incident energies (a) $E_{\nu}=$1
MeV and (b) $E_{\nu}=$ 10 keV, in units of $2c_{V}\rho+\rho^{2}$
and $\rho\equiv\frac{\sqrt{2}\pi}{3}\frac{\alpha}{G_{F}}\braket{\mathbbm{r}_{\nu}^{2}}^{(\mathrm{eff})}$.\label{fig:ds/dT_r2}}
\end{figure}

The differential cross sections due to the interaction with \textbf{$\braket{\mathbbm{r}_{\nu}^{2}}^{(\mathrm{eff})}$},
i.e., by taking $d\sigma^{(\mathrm{h.c.})}/dT-d\sigma^{(w)}/dT$ with
$\mathbbm{q}_{\nu}=0$ in Eq.~(\ref{eq:dS/dT_h.c.}), are given in
Fig.~\ref{fig:ds/dT_r2}. Since the charge radius interaction takes
the same contact form as the weak interaction, it is natural to expect
the failure of the EPA scheme, so not shown in the figure. The main
difference between the charge radius and weak interactions is that
the former depends on the atomic vector-current response, while the
latter on the atomic vector-minus-axial-vector-current (V-A) response.
However, as can be seen from the comparison of Figs.~\ref{fig:ds/dT_r2}
and \ref{fig:ds/dT_weak}, both differential cross sections share
very similar $T$-dependence. The differences between the MCRRPA and
FEA results are also similar to the case of weak scattering.

\subsection{Reactor Antineutrinos \label{sub:TEXONO}}

Existing data from reactor neutrino experiments using germanium ionization
detectors~\cite{Li:2002pn,Wong:2006nx,Beda:2012zz,Beda:2013mta}
provide excellent platform to investigate the atomic ionization effects
induced by neutrino electromagnetic interactions. The sensitivities
depend on the detectable threshold of the differential cross section,
as well as the neutrino flux but are mostly independent to the neutrino
energy. Therefore, the enormous $\bar{\nu}_{e}$ flux (order of $10^{13}\,\mathrm{cm^{-2}\, s^{-1}}$,
at a typical distance of $20\,\mathrm{m}$ from the reactor core)
at the MeV-range energy from nuclear power reactors is a well-suited
source. The germanium detectors, with their excellent energy resolution
and sub-keV threshold, are ideal as means of studying these effects.
The experimental features as peaks or edges at the definite $K$-
and $L$-X-rays energies as well as with predictable intensity ratios
provide potential smoking-gun signatures of these effects~\cite{Wong:2010pb,Chen:2014dsa}.

\begin{figure}
\includegraphics[width=0.8\textwidth]{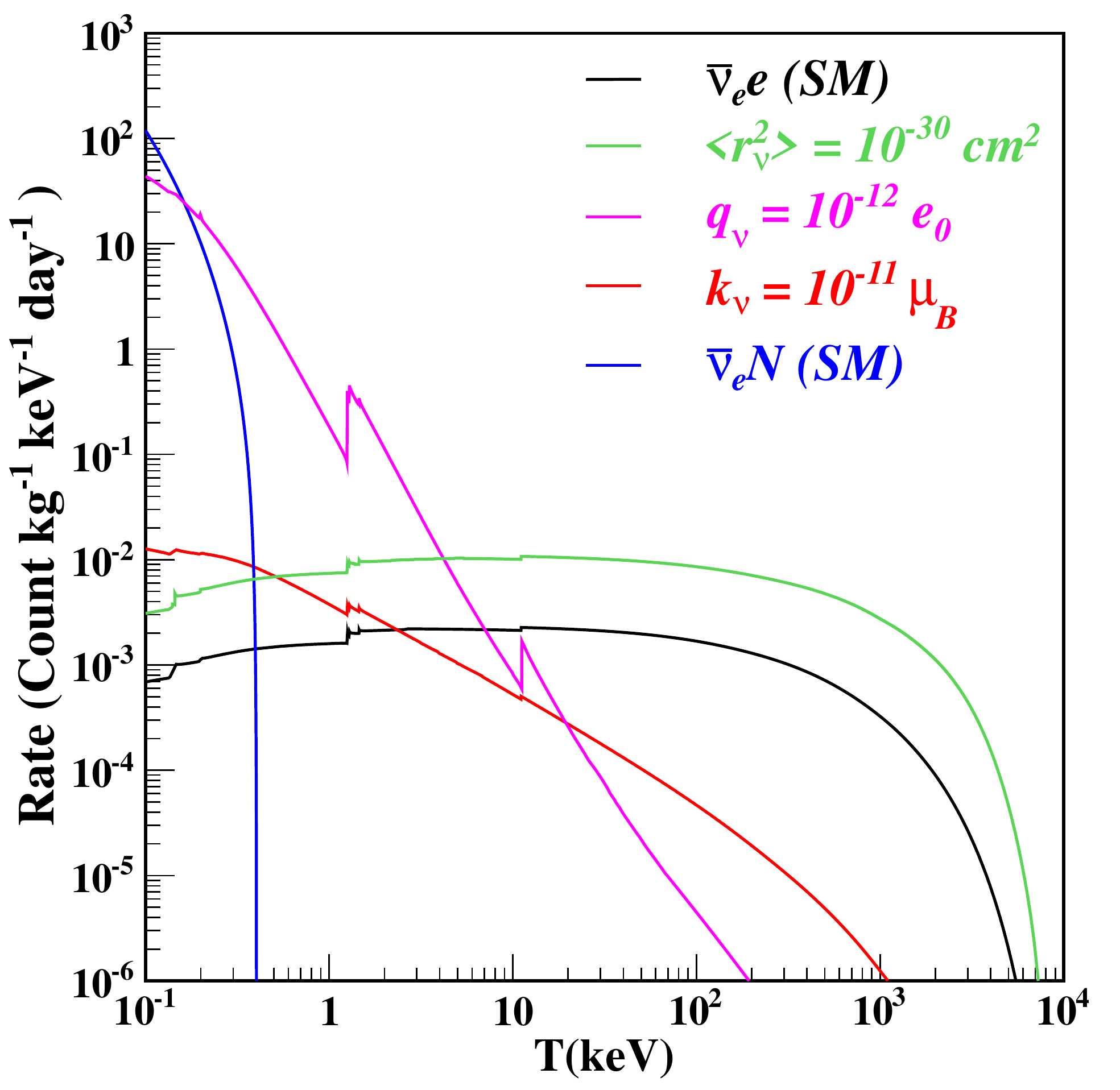} \protect\caption{Expected measurable spectra with Ge on the various neutrino electromagnetic
effects from reactor neutrinos ($\bar{\nu}_{e}$) at a flux of $10^{13}\,\mathrm{cm}^{-2}\,\mathrm{s}^{-1}$.
The spectra from SM weak processes involving the electrons ($\bar{\nu}_{e}e$)
and the nucleus ($\bar{\nu}_{e}N$) are also included for comparisons.\label{fig:all}}
\end{figure}

Denoting the reactor $\bar{\nu}_{e}$ spectrum by $\phi(E_{\nu})$,
the measured differential spectra $\braket{d\sigma/dT}$ is related
to the theoretical formulae of Eqs.~\ref{eq:dS/dT_weak}, \ref{eq:dS/dT_F1}
and \ref{eq:dS/dT_F2}, via: 
\begin{equation}
\left\langle \frac{d\sigma}{dT}\right\rangle =\frac{\int dE_{\nu}\,\phi(E_{\nu})\frac{d\sigma}{dT}(E_{\nu})}{\int dE_{\nu}\,\phi(E_{\nu})}\,.
\end{equation}
The measurable spectra due to weak interactions, neutrino magnetic
moments at \textbf{$\bbkappa_{\nu}^{\mathrm{(eff)}}=10^{-11}\,\mu_{\mathrm{B}}$},
milli-charges at $\mathbbm{q}_{\nu}=10^{-12}\, e$ and charge radius
at $\braket{\mathbbm{r}_{\nu}^{2}}^{(\mathrm{eff})}=[6\times10^{-3}\,\mathrm{fm}]^{2}$
at a reactor $\bar{\nu}_{e}$ flux of $10^{13}\,\mathrm{cm}^{-2}\,\mathrm{s}^{-1}$
are depicted in Fig.~\ref{fig:all}. These are compared with most
sensitive data set from the TEXONO~\cite{Li:2002pn,Wong:2006nx}
and GEMMA~\cite{Beda:2012zz,Beda:2013mta} experiments and the corresponding
limits at 90\% CL are listed in Table~\ref{tab:summary}. Standard
algorithms were adopted to provide best-fit and confidence intervals
to the data (see, for example, the Statistics Section of Ref.~\cite{Agashe:2014kda}).
Also shown are the potential sensitivities of a realistic next-generation
measurements using Ge with sensitivities as low as 100 eV and at a
background level of 1 count/kg-keV-day. 

Both Fig.~\ref{fig:all} and Table~\ref{tab:summary} confirm the
merits of detectors with low-threshold and good energy resolution
in the studies of $\bbkappa_{\nu}^{\mathrm{(eff)}}$ and $\mathbbm{q}_{\nu}$,
where the $d\sigma/dT$ formulae are enhanced as $T\rightarrow0$.
For $\braket{\mathbbm{r}_{\nu}^{2}}^{(\mathrm{eff})}$, detectors
with larger mass like CsI(Tl)~\cite{Deniz:2010mp} making measurements
at the MeV energy range to benefit from the better signal-to-background
ratios would provide better sensitivities. .

\begin{sidewaystable*}
\begin{ruledtabular} %
\begin{tabular}{lcccccc}
Data Set  & Reactor-$\bar{\nu}_{e}$  & Data Strength  & Analysis  & \multicolumn{3}{c}{Bounds at 90\% CL }\tabularnewline
 & Flux  & Reactor ON/OFF  & Threshold  & $\bbkappa_{\bar{\nu}_{e}}^{\mathrm{(eff)}}$  & $\mathbbm{q}_{\bar{\nu}_{e}}$  & $\langle\mathbbm{r}_{\bar{\nu}_{e}}^{2}\rangle^{\mathrm{(eff)}}$ \tabularnewline
 & (${\rm \times10^{13}~cm^{-2}s^{-1}}$)  & (kg-days)  & (keV)  & ($\times10^{-11}~\mu_{{\rm B}}$)  & ($\times10^{-12}$)  & ($\times10^{-30}~\mbox{\ensuremath{\mathrm{cm}}}^{2}$) \tabularnewline
\hline 
TEXONO 187~kg CsI~\cite{Deniz:2010mp}  & 0.64  & 29882.0/7369.0 & 3000  & < 22.0  & < 170  & < 0.033\tabularnewline
TEXONO 1~kg Ge~\cite{Li:2002pn,Wong:2006nx}  & 0.64  & 570.7/127.8  & 12  & < 7.4  & < 8.8  & < 1.40\tabularnewline
GEMMA 1.5~kg Ge~\cite{Beda:2012zz,Beda:2013mta}  & 2.7  & 1133.4/280.4  & 2.8  & < 2.9  & < 1.1  & < 0.80\tabularnewline
TEXONO Point-Contact Ge~\cite{Chen:2013lba,Chen:2014dsa}  & 0.64  & 124.2/70.3  & 0.3  & < 26.0  & < 2.1  & < 3.20\tabularnewline
Projected Point-Contact Ge  & 2.7  & 800/200  & 0.1  & < 1.7  & < 0.06  & < 0.74\tabularnewline
Sensitivity at $1\%$ of SM & --- & --- & --- & $\sim$0.023 & $\sim$0.0004 & $\sim$0.0014\tabularnewline
\end{tabular}\end{ruledtabular} \protect\caption{Summary of experimental limits at 90\% CL on the various neutrino
electromagnetic parameters studied in this work using selected reactor
neutrino data. The projected sensitivities of measurements at the
specified realistically experimental parameters are also shown. The
last row illustrates the effective lower bounds to the sensitivities
when a 1\% measurement of the SM cross-section could be achieved,
at threshold of 0.1 keV for $\bbkappa_{\bar{\nu}_{e}}^{\mathrm{(eff)}}$and
$\mathbbm{q}_{\bar{\nu}_{e}}$, and 3 MeV for $\langle\mathbbm{r}_{\bar{\nu}_{e}}^{2}\rangle^{\mathrm{(eff)}}$,
respectively. \label{tab:summary}}

\label{tab::results} 
\end{sidewaystable*}

\subsection{Neutrinos of Tritium $\beta$ Decay \label{sub:3H_beta}}

\begin{figure}
\includegraphics[width=0.49\textwidth]{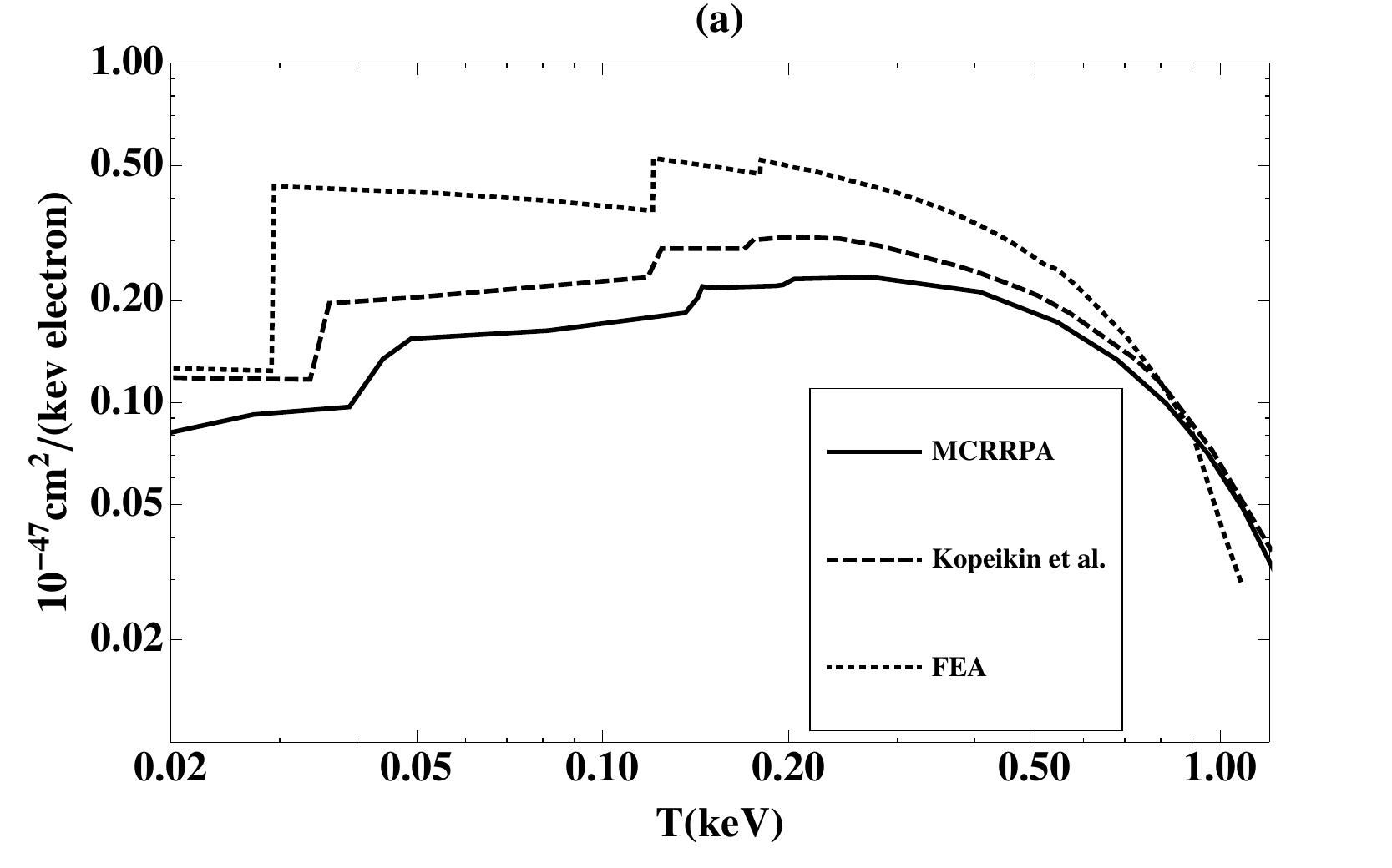}\includegraphics[width=0.49\textwidth]{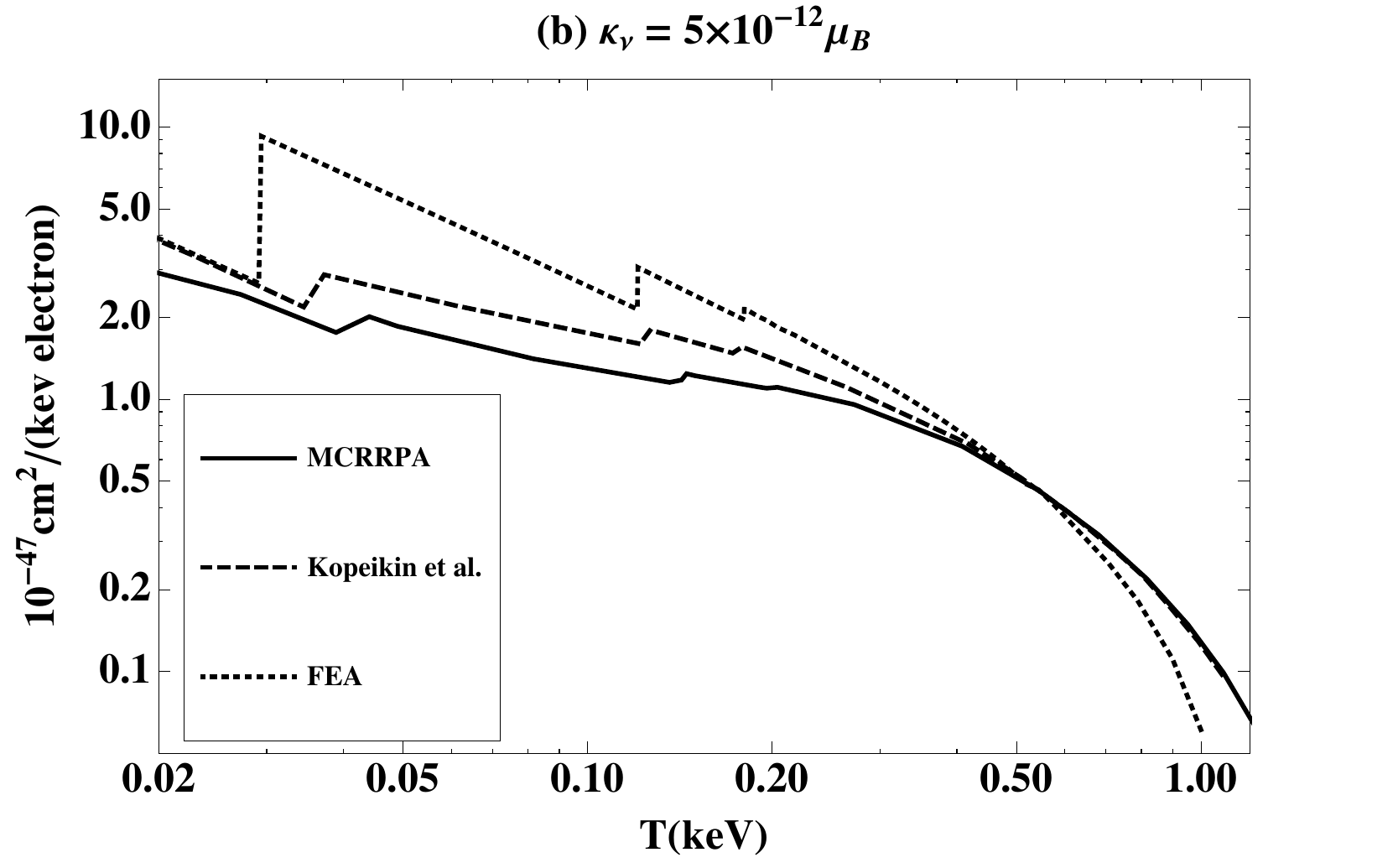}

\protect\caption{Differential cross sections of germanium ionization by neutrinos of
tritium $\beta$ decay through (a) weak and (b) neutrino magnetic
moment interaction assuming $\bbkappa_{\nu}^{\mathrm{(eff)}}=5\times10^{-12}\mu_{\mathrm{B}}$.
\label{fig:ds/dT_NMM_3H}}
\end{figure}

The possibility of using the very low energy $\beta$ neutrinos from
tritium decay to constrain neutrino magnetic moments was discussed
in Refs.~\cite{Kopeikin:2003bx,Giomataris:2003bp,McLaughlin:2003yg}.
In Fig.~\ref{fig:ds/dT_NMM_3H}, we compare the convoluted differential
cross sections calculated by our MCRRPA approach, Ref.~\cite{Kopeikin:2003bx},
and the FEA scheme.

As shown by the figure, below $T<1\,\mathrm{keV}$, FEA predicts larger
cross sections for both neutrino weak and magnetic moment scattering
than the two realistic many-body calculations. This echoes our previous
argument that the Voloshin sum rule and FEA only poses an upper limit
on cross sections, and the binding of an electron is not the only
factor that determines whether FEA can be a good approximation or
not. For $T>0.9\,\mathrm{keV}$ and $T>0.5\,\mathrm{keV}$, FEA predictions
drop quickly below the realistic calculations for weak and magnetic
moment scattering, respectively. This is mainly because the maximum
energy transfer allowed by FEA: $T_{\max}=1.2\,\mathrm{keV}$ ($Q$
value for tritium $\beta$ decay is 18.6 keV) heavily restricts the
allowed final-state phase space for scattering.

While our MCRRPA approach agrees with the previous many-body calculations~\cite{Kopeikin:2003bx}
in the $T>0.9\,\mathrm{keV}$ and $T>0.5\,\mathrm{keV}$ regions for
weak and magnetic moment scattering, respectively; our results are
comparatively smaller at lower $T$. This discrepancy is mostly related
to the treatments in atomic many-body physics: (i) Ref.~\cite{Kopeikin:2003bx}
adopted the same framework as Refs.~\cite{Fayans:1992kk,Kopeikin:1997ge,Fayans:2001pg}
by using the relativistic Dirac-Hartree-Fock method with a local exchange
potential to solve the atomic ground-state structure, while we used
he exact non-local Fock potential. (ii) The local exchange potential
used by Ref.~\cite{Kopeikin:2003bx} is adapted from Ref.\cite{Moruzzi:1978jw}.
This local exchange potential is designed to describe the ground-state
structure of several metals (with Z<50) in the framework of density
functional theory (DFT), therefore, it is not a surprise that it fits
better the $M$-shell single particle energies of germanium crystal
than our atomic calculations, because solid effects have been accounted
for to some extent. (iii) It is known to be challenging to extend
DFT to excited states (such as the ionization states which are relevant
here); it is not clear how well the simplified mean-field scheme used
by Ref.~\cite{Kopeikin:2003bx} can reproduce the photoabsorption
data, say for $T>100\,\mathrm{eV}$\textemdash which we take as a
very important benchmark for the computation of transition matrix
elements.

\section{Summary and Prospects \label{sec:summary}}

In this paper, we show that the multiconfiguration relativistic random
phase approximation provides a good description for the structure
of germanium atoms and the photoabsorption data of germanium solid
at photon energy $\gtrsim80\,\mathrm{eV}$. These benchmark calculations
justify a good understanding of how germanium detectors respond to
neutrinos, through weak and possible electromagnetic interactions,
with a threshold as low as $100\,\mathrm{eV}$.

After taking atomic ionization effects into account, existing reactor
neutrino data with germanium detectors~\cite{Beda:2012zz,Beda:2013mta}
provide the most stringent direct experimental limits on neutrino
millicharge and magnetic moments: $1.1\times10^{-12}\, e$ and $2.9\times10^{-11}\,\mu_{\mathrm{B}}$
at 90\% confidence level, respectively. Future experiments with 100
eV threshold can target at the\textcolor{red}{{} }$10^{-14}\, e$\textcolor{red}{{}
}and $10^{-12}\,\mu_{\mathrm{B}}$ sensitivity range. In particular,
there is substantial enhancement of the millicharge-induced cross
section at low energy, providing smoking-gun signatures for positive
signals. Charge-radius-induced interactions, on the other hand, do
not have enhancement at low energy, such that the best sensitivities
are obtained in experiment~\cite{Deniz:2010mp} with larger detector
mass operating at the MeV energy range where the signal-to-background
ratio is much more favorable. 

The approach explored in this article as well as adopted by current
laboratory experiments and astrophysics studies rely on searching
possible anomalous effects relative to those produced by SM electroweak
processes. It would therefore be experimentally difficult to probe
non-standard effects less than, for example, 1\% that of SM. There
are certain fundamental (limited by physics rather than technology)
lower bounds where such laboratory limits and astrophysics constraints
can reach, as illustrated in Table \ref{tab:summary}. This limitation
can be evaded, at least \textit{conceptually}, by the analog of \textquotedbl{}appearance\textquotedbl{}
experiments with the studies of detector channels where the SM background
vanishes. For instance, in the case of Majorana neutrinos with transition
magnetic moments, one can look for signatures of final-state neutrinos
with a different flavor in a pure and intense neutrino beam which
passes through a dense medium or an intense magnetic field. While
there is no fundamental constraint to the lower reach of the sensitivities,
realistic experiments are still many order-of-magnitude less sensitive
than the reactor neutrino bounds~\cite{GonzalezGarcia:1995fy,Frere:1996gb}.
\begin{acknowledgments}
We acknowledge the support from the NSC/MOST of ROC under Grant Nos.
102-2112-M-002-013-MY3 (JWC, CLW, CPW), 102-2112-M-259-005 and 103-2112-M-259-003
(CPL); the CTS and CASTS of NTU (JWC, CLW, CPW).

\appendix*
\end{acknowledgments}

\section{Multipole Expansion \label{sec:app-1}}

First we set up the coordinate system so that the 3-momentum transfer
by neutrinos is along the $z$-axis, i.e., the Cartesian unit vector
$\hat{e}_{3}=\vec{q}/|\vec{q}|$. The transformation between the unit
vectors in the spherical ($\hat{\epsilon}^{\lambda=\pm1,0}$) and
Cartesian ($\hat{e}^{i=1,2,3}$) systems is then given by 
\begin{equation}
\hat{\epsilon}^{\pm1}=\mp\frac{1}{\sqrt{2}}(\hat{e}^{1}\pm i\hat{e}^{2})\,,\qquad\hat{\epsilon}^{0}=\hat{e}^{3}\,.
\end{equation}
The spherical component of a vector $\vec{V}$, denoted by $\lambda$,~%
\footnote{$V^{\lambda=0}$ should not to be confused with the time component
of a Lorentz 4-vector.%
} is 
\begin{equation}
V^{\lambda}=\hat{\epsilon}^{\lambda}\cdot\vec{V}\,.
\end{equation}

According to Eq.~(\ref{eq:MCRRPA-scat.amp.}), the perturbing field
that gives rise to atomic ionization by neutrino electromagnetic interactions
takes the form 
\begin{align}
\left<\Psi_{f}\left|v_{+}^{(\gamma)}\right|\Psi_{i}\right> & =\frac{4\pi\alpha}{q^{2}}\left\{ j_{0}^{(\gamma)}\left\langle \Psi_{f}\left|\int d^{3}x\, e^{i\vec{q}\cdot\vec{x}}\hat{\mathcal{J}^{0}}(\vec{x})\right|\Psi_{i}\right\rangle \right.\nonumber \\
 & \left.+\sum_{\lambda=\pm1,0}(-1)^{\lambda}j_{\lambda}^{(\gamma)}\left\langle \Psi_{f}\left|\int d^{3}x\, e^{i\vec{q}\cdot\vec{x}}\hat{\epsilon}^{-\lambda}\cdot\hat{\vec{\mathcal{J}}}(\vec{x})\right|\Psi_{i}\right\rangle \right\} \,.
\end{align}
Using the relations 
\begin{align}
e^{i\vec{q}\cdot\vec{x}} & =\sum_{J=0}^{\infty}\sqrt{4\pi(2J+1)}i^{J}j_{J}(\kappa r)Y_{J}^{0}(\Omega_{x})\,,\\
e^{i\vec{q}\cdot\vec{x}}\hat{\epsilon}^{0} & =\frac{-i}{\kappa}\sum_{J=0}^{\infty}\sqrt{4\pi(2J+1)}i^{J}\overrightarrow{\nabla}\left[j_{J}(\kappa r)Y_{J}^{0}(\Omega_{x})\right]\,,\\
e^{i\vec{q}\cdot\vec{x}}\hat{\epsilon}^{\pm1} & =-\sum_{J\geq1}\sqrt{2\pi(2J+1)}i^{J}\left\{ \frac{1}{\kappa}\overrightarrow{\nabla}\times\left[j_{J}(\kappa r)\mathcal{Y}_{JJ1}^{\pm1}(\Omega_{x})\right]\pm j_{J}(\kappa r)\mathcal{Y}_{JJ1}^{\pm1}(\Omega_{x})\right\} \,,
\end{align}
where $|\vec{q}|\equiv\kappa$, $|\vec{x}|\equiv r$, $j_{J}(\kappa r)$
is the spherical Bessel function of order $J$, $Y_{J}^{M}(\Omega_{x})$
the spherical harmonics, and $\mathcal{Y}_{Jl1}^{M}(\Omega_{x})$
the vector spherical harmonics formed by adding $Y_{l}^{m}(\Omega_{x})$
and $\hat{\epsilon}^{\lambda}$ to be an angular momentum eigenstate
$\ket{JM}$: 
\begin{equation}
\mathcal{Y}_{Jl1}^{M}(\Omega_{x})\equiv\sum_{m\lambda}\left\langle lm1\lambda|l1JM\right\rangle Y_{l}^{m}(\Omega_{x})\hat{\epsilon}^{\lambda}\,,
\end{equation}
the perturbing field is expanded as 
\begin{align}
\left<\Psi_{f}\left|v_{+}^{(\gamma)}\right|\Psi_{i}\right> & =\frac{4\pi\alpha}{q^{2}}\left\{ \sum_{J=0}^{\infty}\sqrt{4\pi(2J+1)}i^{J}\left[j_{0}^{(\gamma)}\braket{\hat{C}_{J0}(\kappa)}-j_{3}^{(\gamma)}\braket{\hat{L}_{J0}(\kappa)}\right]\right.\nonumber \\
 & \left.+\sum_{J\ge1}^{\infty}\sqrt{2\pi(2J+1)}i^{J}\sum_{\lambda=\pm1}j_{\lambda}^{(\gamma)}\left[\braket{\hat{E}_{J-\lambda}(\kappa)}-\lambda\braket{\hat{M}_{J-\lambda}(\kappa)}\right]\right\} \,.
\end{align}
The various spherical multipole operators are defined by 

\begin{eqnarray}
\hat{C}_{JM}(\kappa) & = & \int d^{3}x\,[j_{J}(kr)Y_{JM}]\,\hat{\mathcal{J}}^{0}(\vec{x})\,,\\
\hat{L}_{JM}(\kappa) & = & \frac{i}{\kappa}\int d^{3}x\,\overrightarrow{\nabla}[j_{J}(\kappa r)Y_{JM}(\Omega_{x})]\cdot\hat{\vec{\mathcal{J}}}(\vec{x})\\
\hat{E}_{JM}(k) & = & \frac{1}{\kappa}\int d^{3}x\,\overrightarrow{\nabla}\times[j_{J}(\kappa r)\mathcal{Y}_{JJ1}^{M}(\Omega_{x})]\cdot\hat{\vec{\mathcal{J}}}(\vec{x})\\
\hat{M}_{JM}(k) & = & \int d^{3}x\,[j_{J}(\kappa r)\mathcal{Y}_{JJ1}^{M}(\Omega_{x})]\cdot\hat{\vec{\mathcal{J}}}(\vec{x})\,.
\end{eqnarray}
Each operator has its specific angular momentum and parity selections
rules that restrict the possible initial-to-final-state transitions. 

When dealing with weak interactions, the axial vector current operator
$\hat{\mathcal{J}}_{5}(\vec{x})$ generates additional four types
of multipole operators $\hat{C}_{JM}^{5}$, $\hat{L}_{JM}^{5}$, $\hat{E}_{JM}^{5}$,
and $\hat{M}_{JM}^{5}$. They are obtained simply by replacing the
vector current operator $\hat{\mathcal{J}}(\vec{x})$ with $\hat{\mathcal{J}}_{5}(\vec{x})$
in the above definitions. 

\bibliographystyle{apsrev4-1}
\bibliography{Ge_EM_full}

\end{document}